\documentclass[aps,prd,reprint,superscriptaddress,amsmath,amssymb,nofootinbib,floatfix]{revtex4-2}
\usepackage{graphicx}
\usepackage{dcolumn}
\usepackage[utf8]{inputenc}
\usepackage{xcolor}
\usepackage[unicode=true,colorlinks,linkcolor=blue,anchorcolor=blue,urlcolor=blue,citecolor=blue,breaklinks=true]{hyperref}

\newcommand{\ii}{\mathrm{i}\,}

\newcommand{\pararrow}{\mathord{\buildrel{\lower3pt\hbox{$\scriptscriptstyle\leftrightarrow$}}\over {\partial}}} 
\newcommand{\pararrowk}[1]{\mathord{\buildrel{\lower3pt\hbox{$\scriptscriptstyle\leftrightarrow$}}\over {\partial}\hspace*{-0.18em}{}^#1}\hspace*{-0.18em} \,} 
\newcommand{\mytrace}[1]{\langle #1 \rangle} 

\usepackage{physics}
\usepackage{bm}

\newcommand{\qfnu}{\affiliation{College of Physics and Engineering, Qufu Normal University, Qufu 273165, China}}

\newcommand{\itp}{\affiliation{CAS Key Laboratory of Theoretical Physics, Institute of Theoretical Physics, Chinese Academy of Sciences, Beijing 100190, China}}

\newcommand{\imp}{\affiliation{Institute of Modern Physics, Chinese Academy of Sciences, Lanzhou 730000, China}}

\newcommand{\snst}{\affiliation{School of Nuclear Science and Technology, University of Chinese Academy of Sciences, Beijing 101408, China}}

\newcommand{\scnt}{\affiliation{Southern Center for Nuclear-Science Theory (SCNT), Institute of Modern Physics, Chinese Academy of Sciences, Huizhou 516000, Guangdong, China}}

\newcommand{\tju}{\affiliation{Center for Joint Quantum Studies and Department of Physics, School of Science, Tianjin University, Tianjin 300350, China}}

\begin{document}
	
	\title{Pionic transitions of the spin-2 partner of $X(3872)$ to $\chi_{cJ}$}
	
	\author{Shi-Dong Liu} \qfnu
	\author{Fan Wang} \qfnu
	\author{Zhao-Sai Jia} \qfnu\itp
	\author{Gang Li}\email{gli@qfnu.edu.cn} \qfnu
	\author{Xiao-Hai Liu}\email{xiaohai.liu@tju.edu.cn}\tju
	\author{Ju-Jun Xie} \email{xiejujun@impcas.ac.cn} \imp \snst \scnt

\begin{abstract}
We investigated the pionic transitions between the $X_2$ [spin-2 partner of the $X(3872)$] and $\chi_{c1,2}$ using a nonrelativistic effective field theory. The $X_2$ is assumed to be a bound state of the $D^{*}$ and $\bar{D}^*$ mesons and to decay through several kinds of loops, including the bubble, triangle and box loops. Within the present model, the widths for the single-pion decays $X_2\to\pi^0\chi_{cJ}$ are predicted to be about $3$--$30$ keV. For the dipion decays, the widths are a few keVs. These widths yield a branching fraction of $10^{-3}$--$10^{-2}$. The ratio $R_{\mathrm{c}0}=\Gamma (X_2\to\pi^+\pi^-\chi_{cJ})/\Gamma (X_2\to\pi^0\pi^0\chi_{cJ}) \simeq 1.6$, which is a bit smaller than the expected value of $2$, and $R_{21}=\Gamma (X_2\to\pi\pi\chi_{c2})/\Gamma (X_2\to\pi\pi\chi_{c1}) \simeq 0.85$. These ratios are nearly independent of the $X_2$ mass and the coupling constants, which might be a good quantity for the experiments. Moreover, the invariant mass spectra of the $\pi^0\chi_{cJ}$ final state for the dipion processes are presented, showing a cusp structure at the $D {\bar D}^*$ threshold enhanced and narrowed by the nearby triangle singularity.
		
\end{abstract}

\date{\today}

\maketitle
\section{Introduction}\label{sec:intro}

After the discovery of the $X(3872)$ state by the Belle collaboration in 2003 \cite{choi2003PRL91-262001}, the exotic states that do not fit into the classical quark model scheme have received considerable attention during the past two decades. The unusual properties of the exotic states challenge the conventional quark model interpretations, but also provide a different perspective on strong interactions. So far a lot of exotic states have been observed experimentally, for instance, the $Z_c (3900)$ \cite{ablikim2013PRL110-252001,liu2013PRL110-252002,abazov2018PRD98-052010} and $Z_{cs}(3985)$ \cite{ablikim2021PRL126-102001} in the charmonium sector and the $Z_b(10610)$ and $Z_b(10650)$ \cite{bondar2012PRL108-122001} in the bottomonium sector. Various theoretical approaches, including phenomenological extensions of the quark model, non-relativistic effective field theory, and lattice QCD calculations, have be used to understand the nature of exotic states. More detailed experimental and theoretical studies on the exotic states can be found in recent reviews \cite{chen2016PR639-1,lebed2017PPNP93-143,guo2018RMP90-015004,brambilla2020PR873-1,kalashnikova2019P62-568,meng2023PR1019-2266}.

The $X(3872)$ was first observed in the decay processes $B^{\pm}\to K^{\pm}\pi^+\pi^- J/\psi$ as a narrow peak in the invariant mass distribution of the $\pi^+\pi^- J/\psi $ final state \cite{choi2003PRL91-262001}, and subsequently confirmed by various experiments \cite{abazov2004PRL93-162002,acosta2004PRL93-072001,aubert2005PRD71-071103}. The most up-to-date mass of the $X(3872)$ is $(3871.65\pm 0.06)~\mathrm{MeV}$, quite close to the $D^{*0}\bar{D}^0$ threshold ($3871.69~\mathrm{MeV}$) ~\cite{workman2022PTEP2022-083C01}. Its quantum numbers 
was determined to be $J^{PC}=1^{++}$ \cite{aaij2013PRL110-222001,aaij2015PRD92-011102}. In view of the threshold proximity of the $X(3872)$, the molecular picture becomes a popular interpretation \cite{fleming2008PRD78-094019,dubynskiy2008PRD77-014013,guo2013PLB725-127,wang2014EPJC74-2891,guo2015PLB742-394,mehen2015PRD92-034019,guo2018RMP90-015004,kalashnikova2019P62-568,dai2020PRD101-054024,wang2022PRD106-074015,wu2021EPJC81-193,meng2021PRD104-094003}. However, other interpretations, including the compact tetraquark \cite{chen2022CTP74-025201,shi2021PRD103-094038,maiani2014PRD89-114010,barnea2006PRD73-054004,maiani2005PRD71-014028,wang2024PRD109-014017,wang2014PRD89-054019}, a hybrid charmonium state
with constituents $c\bar{c}g$ \cite{close2003PLB574-210,li2005PLB605-306}, and a conventional charmonium state \cite{barnes2004PRD69-054008,quigg2005NPBPS142-87,eichten2004PRD69-094019,suzuki2005PRD72-114013,achasov2024PRD109-036028,achasov2024x-} also appear feasible. Although the $X(3872)$ is the most well-studied exotic state either from the experimental or theoretical points of view \cite{brambilla2020PR873-1,kalashnikova2019P62-568} (and references therein), the consensus about its nature is still not achieved.

To deepen understanding of the $X(3872)$ nature, the direct way is to search for more production and decay modes. Apart from that, investigating its analogues is also helpful. Based on the heavy quark spin symmetry, there might exist a series of spin partners of the $X(3872)$ with the configuration of charmed-anticharmed mesons in the charmonium sector. The masses and quantum numbers of these partners have been theoretically predicted in Refs.~\cite{nieves2012PRD86-056004,peng2023PRD108-114001,guo2013PRD88-054007,tornqvist2004ax-,tornqvist1994ZPCPF61-525,mutuk2018EPJC78-904,wang2021IJMPA36-2150107,wang2020PRD102-014018}. Searching experimentally for these partners appears to make progress in 2022 when the Belle Collaboration observed a structure in the invariant mass distribution of $\gamma\psi(2S)$ with a mass of $(4014.3\pm 4.0\pm 1.5)~\mathrm{MeV}$ and a width of $(4\pm 11\pm 6)~\mathrm{MeV}$ \cite{wang2022PRD105-112011}. This structure lies fairly near the $D^{*0}\bar{D}^{*0}$ threshold, with a tiny mass difference $ \Delta m = 4014.3 -2 m_{D^{*0}}=0.6~\mathrm{MeV}$. Moreover, the measured mass and width agree well with the theoretically predicted mass of ($4012$--$4015)$~$\mathrm{MeV}$~\cite{nieves2012PRD86-056004,peng2023PRD108-114001,guo2013PRD88-054007,tornqvist2004ax-,tornqvist1994ZPCPF61-525} and width of ($2$--$8$)~$\mathrm{MeV}$~\cite{albaladejo2015EPJC75-547} of the $2^{++}$ $D^*\bar{D}^*$ partner of the $X(3872)$. Hence, this structure could be a good candidate for the spin-2 partner of the $X(3872)$, called the $X_2 $ hereafter. Recently, Shi \textit{et al.}, under the assumption that the $X_2$ is a $D^*\bar{D}^*$ molecule, investigated its radiative transitions to the $J/\psi$ and $\psi(2S)$. It is found that the ratio $\Gamma (X_2\to\gamma J/\psi)/\Gamma (X_2\to\gamma \psi(2S))$ is about 1.5 times larger than the case for the $X(3872)$~\cite{shi2023PLB843-137987}. In Ref.~\cite{zheng2024PRD109-014027}, the partial widths of the $X_2$ decaying into $\omega (\rho^0)J/\psi$ and $\pi^0(\eta)\eta_c$ were predicted. Besides, the production of the $X_2$ in the $e^+e^-$ collisions \cite{shi2024CPL41-031301}, in the $B$ decays \cite{wu2024EPJC84-147}, and in the $Y(4360)$ decays \cite{liu2024a[x-} were also discussed.

To determine the nature of the $X(3872)$, we need compare the predictions from different models with the experimental measurements. However, different theoretical interpretations sometimes give similar results, which leads to difficult understanding of its nature. For instance, the conventional charmonium and molecule models, which provide completely different internal structures of the $X(3872)$, could both interpret experimental measurements $\mathcal{B}(X(3872)\to\pi^0\chi_{c1}) = (3.4\pm 1.6)\times 10^{-2}$ \cite{workman2022PTEP2022-083C01} via the intermediate mesonic loops \cite{wu2021EPJC81-193,achasov2024PRD109-036028}. Together with the theoretical prediction $\mathcal{B}(X(3872)\to\pi^+\pi^-\chi_{c1}) = (0.86 \sim 1.77)\times 10^{-4}$ based on the conventional charmonium picture \cite{achasov2024x-}, it yields
		\begin{equation*}
			\frac{\Gamma(X(3872)\to\pi^+\pi^-\chi_{c1})}{\Gamma(X(3872)\to\pi^0\chi_{c1})}\approx (2.5\sim 5.2)\times 10^{-3}\,,
		\end{equation*} 
		whereas the molecule model also gives the similar ratio of $\mathcal{O}(10^{-3})$ \cite{fleming2008PRD78-094019}. These two theoretical predictions are in agreement with the experimental measurements \cite{ablikim2019PRL122-202001,ablikim2024PRD109-L071101,workman2022PTEP2022-083C01}
		\begin{equation*}
			\frac{\Gamma(X(3872)\to\pi^+\pi^-\chi_{cJ})}{\Gamma(X(3872)\to\pi^0\chi_{cJ})}< 0.2\,.
		\end{equation*}
In this work, we shall, using a nonrelativistic effective field theory, investigate the single-pion and dipion transitions  $X_2\to\pi^0\chi_{c1,2}$ and $X_2\to\pi\pi\chi_{c1,2}$ via intermediate meson loops, including the bubble, triangle, and box diagrams. We first give the Lagrangians in Sec. \ref{sec:lags}, then in Sec. \ref{sec:results} present the calculated results and discussion, and finally give a summary in Sec. \ref{sec:summary}.

\section{Theoretical Consideration}\label{sec:lags}

\begin{figure*}[htbp]
	\centering
	\includegraphics[width=0.8\linewidth]{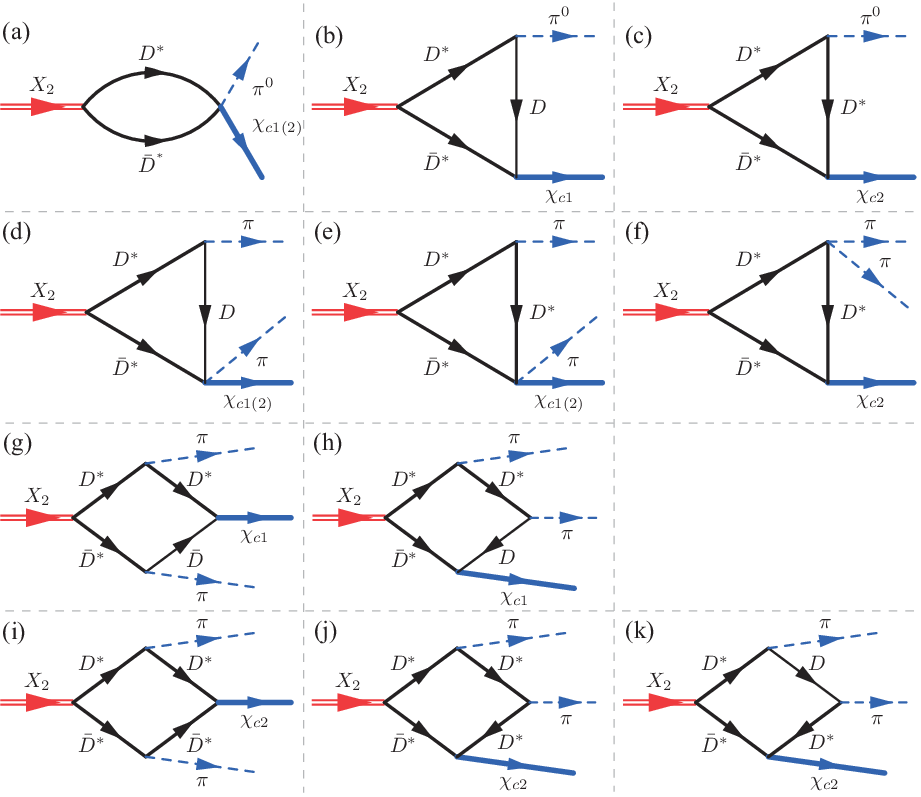}
	\caption{Feynman diagrams for the $X_2\to\pi^0\chi_{cJ}$ and $ X_2\to\pi\pi\chi_{cJ}$ decays via the intermediate meson loops. The charge-conjugated loops are not shown here, but included in our calculations.}
	\label{fig:feyndiags}
\end{figure*}

We assume that the processes of $X_2$ decaying into $\pi^0\chi_{cJ}$ or $\pi\pi\chi_{cJ}$ to occur via the intermediate meson loops shown in Fig. \ref{fig:feyndiags}. The involved interactions of the external particles with the intermediate mesons are described by the effective Lagrangians. In the molecular picture, the $X_2$ is a bound state of the $D^*$ and $\bar{D}^*$ with the following configuration \cite{tornqvist1994ZPC61-525,nieves2012PRD86-056004,hidalgo-duque2013PRD87-076006,guo2013PRD88-054007,baru2016PLB763-20}
\begin{equation}
	\ket*{X_2}=\cos\theta \ket*{D^{*0}\bar{D}^{*0}} + \sin\theta\ket*{D^{*+}D^{*-}}\,,
\end{equation}
where $\theta$ is introduced to describe the proportion of the neutral and charged components. The interaction between the $X_2$ and the $D^*\bar{D}^*$ pairs reads \cite{shi2023PLB843-137987,zheng2024PRD109-014027}
\begin{align}\label{eq:Lx2}
	\mathcal{L}_{X_2} &= X_{2ij} \big(g_X^\mathrm{n}\cos\theta  D^{*0\dagger i}\bar{D}^{*0\dagger j} \nonumber\\
	&+ g_X^\mathrm{c}\sin\theta D^{*+\dagger i}D^{*-\dagger j}\big) + \mathrm{H.c.}\,,
\end{align}
where $g_X^\mathrm{n}$ and $g_X^\mathrm{c}$ are the couplings to the neutral and charged charmed mesons, respectively. Their values can be estimated by the following expression
\begin{equation}\label{eq:coupgx2}
	g_X = \Bigg(\frac{16}{\pi}\sqrt{\frac{2E_B}{\mu}}\Bigg)^{1/2}\,.
\end{equation}
Here $E_B$ is the binding energy relative to the $D^*\bar{D}^*$ threshold and $\mu$ stands for the reduced mass of the $D^*$ and $\bar{D}^*$. For the neutral case of $D^{*0}\bar{D}^{*0}$, $g_X^\mathrm{n}=1.32~\mathrm{GeV^{-1/2}}$, whereas $g_X^\mathrm{c}=2.36~\mathrm{GeV^{-1/2}}$ for the $D^{*+}D^{*-} $ when we take $m_{X_2}=4014.3~\mathrm{MeV}$.

In terms of the heavy quark symmetry and chiral symmetry, the leading order effective Lagrangian for the light pseudoscalar mesons with the heavy charmed mesons is \cite{fleming2008PRD78-094019,mehen2015PRD92-034019,jia2024PRD109-034017,hu2006PRD73-054003,casalbuoni1997PR281-145}
\begin{align}\label{eq:Lagpi}
		& \mathcal{L}_\pi = -g \mytrace{H_a^\dagger H_b \vec{\sigma}\cdot \vec{A}_{ba}} = \nonumber\\
		& \frac{2g}{f_\pi} \big(\ii \epsilon^{ijk} V_a^{\dagger i}V_b^j \partial^k \phi_{ba} + V_a^{\dagger i}P_b \partial^i \phi_{ba}+ P_a^\dagger V_b^i\partial^i\phi_{ba} \big)\,,
\end{align}
where the symbol $\mytrace{\cdots}$ represents the trace and $H_a \equiv \vec{V}_a\cdot\vec{\sigma}+P_a$ is the heavy charmed mesons in the two-component notation. Explicitly, $V_a \equiv (D^{*0},\,D^{*+})$ and $P_a=(D^0,\,D^+)$, denoting the vector and pseudoscalar mesons, respectively. $\vec{A}_{ba} = - \grad{\phi_{ba}}/f_\pi$ is the axial current with the Goldstone boson fields
\begin{equation}
	\phi_{ba} = \mqty(\dfrac{1}{\sqrt{2}}\pi^0 & \pi^+\\ \pi^- & -\dfrac{1}{\sqrt{2}}\pi^0)\,,
\end{equation}
and the pion decay constant $f_\pi=132~\mathrm{MeV}$. The axial coupling constant $g$ is taken to be $0.59$~\cite{wu2021EPJC81-193}.


The coupling of the $P$-wave charmonia $\chi_{cJ}$ to a pair of charmed and anticharmed mesons is described as~\cite{fleming2008PRD78-094019,guo2011PRD83-034013,mehen2015PRD92-034019}
\begin{align}\label{eq:LagchiDD}
	\mathcal{L}_{\chi DD} &= \ii \frac{g_1}{2} \mytrace{\chi^{\dagger i}H_a\sigma^i\bar{H}_a}+\mathrm{H.c.}\nonumber\\
	&=-\ii 2g_1\chi_{c2}^{\dagger ij}V_a^j\bar{V}_a^i - \sqrt{2}g_1\chi_{c1}^{\dagger i} (V_a^i \bar{P}_a + P_a\bar{V}_a^i) \nonumber\\
	&+ \ii \frac{1}{\sqrt{3}}g_1 \chi_{c0}^\dagger (V_a^i\bar{V}_a^j + 3P_a\bar{P}_a)+\mathrm{H.c.}
\end{align}
Here $\chi^i$ represents the $\chi_{cJ}$ fields in the form of
\begin{equation}
	\chi^i = \sigma^j \Big(\chi_{c2}^{ij} + \frac{1}{2}\epsilon^{ijk}\chi_{c1}^k+\frac{\delta^{ij}}{\sqrt{3}}\chi_{c0}\Big)\,.
\end{equation}
The $\chi_{c2}^{ij}$ is a symmetric and traceless tensor \cite{guo2011PRD83-034013}. Moreover, $\bar{H}_a = -\vec{\bar{V}}_a\cdot\vec{\sigma}+\bar{P}_a$ is the field for antimesons. Here, the coupling constant $g_1$ is taken to be \cite{mehen2015PRD92-034019}
\begin{equation}
	g_1 = - \sqrt{\frac{m_{\chi_{c0}}}{6}} \frac{1}{f_{\chi_{c0}}},
\end{equation}
with $m_{\chi_{c0}}$ and $f_{\chi_{c0}}=510~\mathrm{MeV}$ being the mass and decay constant of the charmonium state $\chi_{c0}$, respectively. Using the current world averaged mass $m_{\chi_{c0}}=3417.71~\mathrm{MeV}$ \cite{workman2022PTEP2022-083C01}, $g_1$ is about $-1.48~\mathrm{GeV^{-1/2}}$. 

To calculate the loops in Figs. \ref{fig:feyndiags}(a), (d), and (e), we need the Lagrangians for the four-body interactions. The effective Lagrangian for the $\chi_{cJ}$ to the charmed-anticharmed meson pair plus one pion is written as \cite{mehen2015PRD92-034019,fleming2008PRD78-094019}
\begin{align}\label{eq:LchiDDpi}
	\mathcal{L}_{\chi D D \pi} &= \frac{c_1}{2}\mytrace{\chi^{\dagger i} H_a \sigma^j \bar{H}_b}\epsilon_{ijk}A_{ab}^k + \mathrm{H.c.}\nonumber\\
	& = -c_1 \chi_{c2}^{\dagger il} (V_a^l\bar{V}_b^j + V_a^j\bar{V}_b^l)\epsilon_{ijk}A_{ab}^k \nonumber\\
	&+ \ii c_1 \chi_{c2}^{\dagger ik} (V_a^i \bar{P}_b+P_a\bar{V}_b^i)A_{ab}^k\nonumber\\
	&+ \frac{c_1}{\sqrt{2}}\chi_{c1}^{\dagger i} (V_a^{j}\bar{V}_b^i + V_a^i\bar{V}_b^j)A_{ab}^j \nonumber\\
	&+ \ii \frac{c_1}{\sqrt{2}} \epsilon_{ijk}\chi_{c1}^{\dagger i}(V_a^j\bar{P}_b+P_a\bar{V}_b^j)A_{ab}^k \nonumber\\
	&+ \sqrt{2}c_1 \chi_{c1}^{\dagger i}P_a\bar{P}_bA_{ab}^{i}\nonumber\\
	&-\ii\frac{2}{\sqrt{3}}c_1\chi_{c0}(V_a^i\bar{P}_b+P_a\bar{V}_b^i)A^i_{ab}+\mathrm{H.c.}\,,
\end{align}
where $c_1$ is an unknown coupling constant. According to the power counting in Appendix \ref{app:pc}, we use the value of $g_1$ to scale $c_1$. The interaction between two charmed mesons and two pions reads
\begin{align}\label{eq:LagDDpipi}
	\mathcal{L}_{DD\pi\pi} &=\mytrace{H_a^\dagger (\ii D_0)_{ba}H_b} \nonumber\\
	&= -\ii 2 V_a^{\dagger i}(V_0)_{ba}V_b^i - \ii 2 P_a^\dagger (V_0)_{ba}P_b\,,
\end{align}
where $D_0=\partial_0-V_0$ and $V_0 = \frac{1}{2} (\xi^\dagger\partial_0\xi+\xi\partial_0\xi^\dagger)$ with $\xi = \exp(\ii \phi/f_\pi)$ \cite{casalbuoni1997PR281-145,jia2024PRD109-034017}.

\section{Numerical Results and Discussion}\label{sec:results}

\subsection{Partial decay widths of $X_2\to\pi^0\chi_{c1,2}$}

Note that for the isospin broken processes $X_2\to\pi^0\chi_{c1,2}$, the contribution from the bubble loop in Fig. \ref{fig:feyndiags}(a) is cutoff($\Lambda$)-dependent [see Eq. \eqref{eq:apptwoloop}] as well as the phase-angle($\theta$)-dependent [see Eq. \eqref{eq:Lx2}]. Moreover, the coupling constant $c_1$ in Eq. \eqref{eq:LchiDDpi} is not well known. Hence, it is better to discuss firstly the contribution from the triangle loops depicted by Figs. \ref{fig:feyndiags}(b) and (c), for which only the phase angle $\theta$ is free. The loop integral functions we used are presented in Appendix \ref{app:loopfunc}.

In Fig.~\ref{fig:threepoint2btres}, the partial decay widths for the $X_2\to \pi^0\chi_{c1,2}$ processes, with contributions only from the triangle loops in Fig. \ref{fig:feyndiags}(b) and (c), are plotted as a function of the phase angle $\theta$. It is seen that the results are sensitive to the phase angle. Near $\theta \approx 30^\circ$, the widths exhibit a valley. With going away from $ \theta \approx 30^\circ $, the widths increase rapidly, and even exceeds the total width of the $X_2$, which is not allowed physically. 

\begin{figure}
	\centering
	\includegraphics[width=0.92\linewidth]{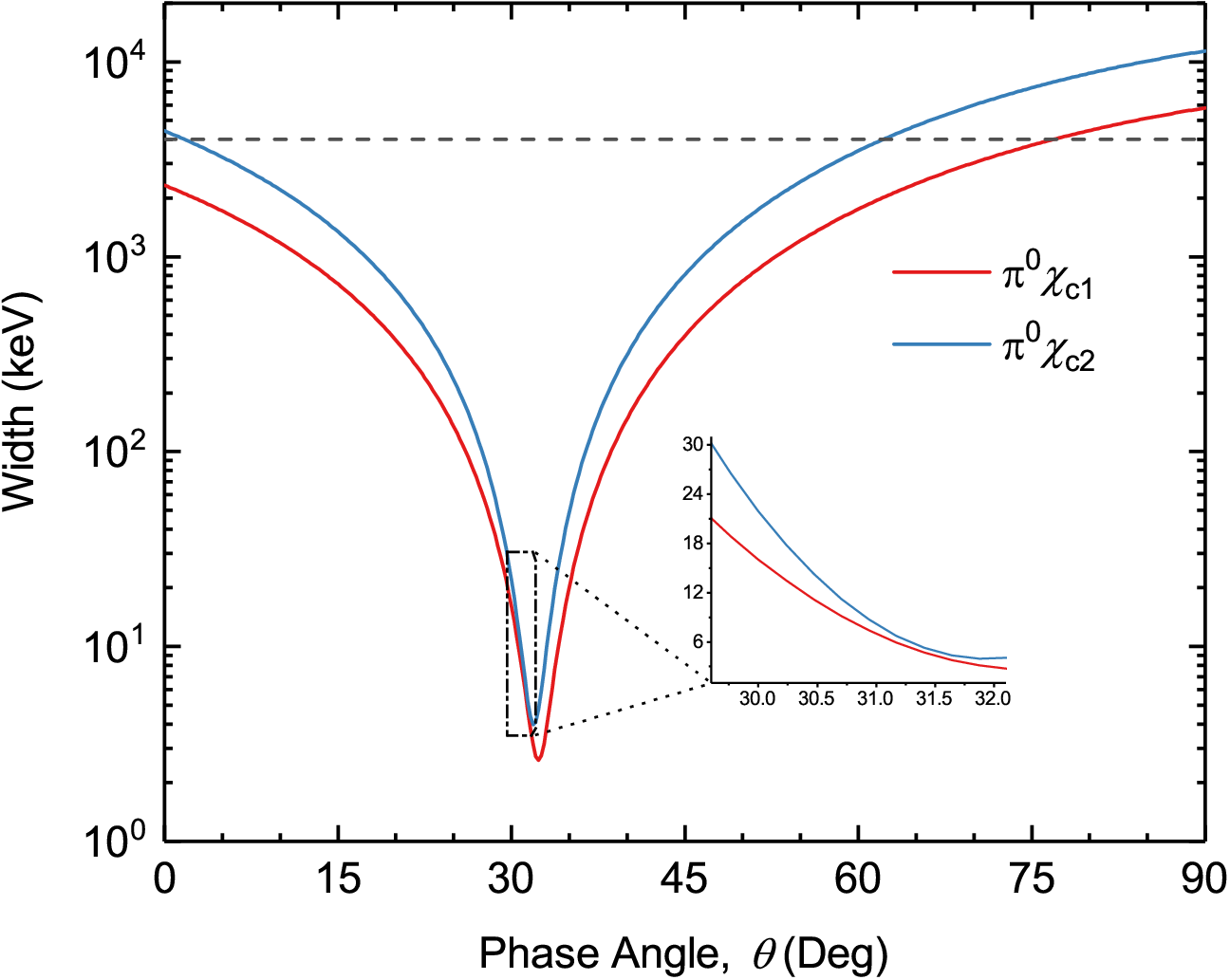}
	\caption{Phase-angle dependence of partial widths for the $X_2\to\pi^0\chi_{c1,2}$ via the triangle loops in Figs. \ref{fig:feyndiags}(b) and (c). The horizontal dashed line indicates the total width of the $X_2$ candidate, $\Gamma_\mathrm{t}(X_2) = (4\pm 11\pm 6)~\mathrm{MeV}$, measured by Belle collaboration \cite{wang2022PRD105-112011}.}
	\label{fig:threepoint2btres}
\end{figure}

\begin{figure}
	\centering
	\includegraphics[width=0.92\linewidth]{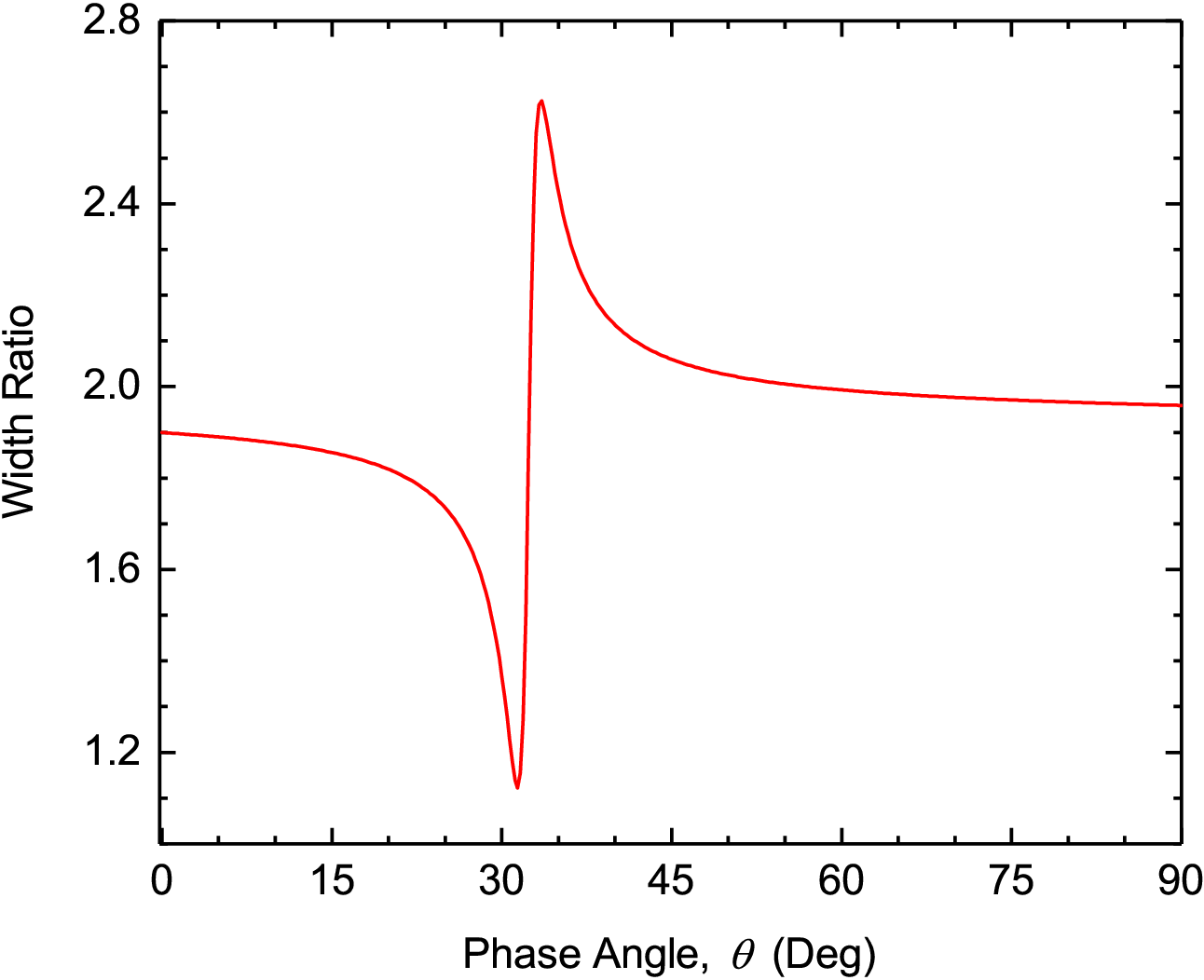}
	\caption{Width ratio of $\Gamma(X_2\to\pi^0\chi_{c2})/\Gamma(X_2\to\pi^0\chi_{c1})$ as a function of the phase angle $\theta$. Only the triangle loops were considered in this calculation.}
	\label{fig:threepoint2ratio}
\end{figure}

In Refs.~\cite{mehen2015PRD92-034019,wu2021EPJC81-193}, the phase angle for the $X(3872)$ was determined to be $20^\circ$--$24^\circ$. To constrain the phase angle for the $X_2$, we assume that the $X_2$ has similar decay modes to those of the $X(3872)$. 
It is found that the width of the $X(3872)\to\pi^0\chi_{c1} $ seems to be equal to that of the $X(3872)\to\pi^0\chi_{c2}$ \cite{workman2022PTEP2022-083C01,mehen2015PRD92-034019,fleming2008PRD78-094019,wu2021EPJC81-193,zhou2019PRD100-094025}. In particular, when considering only the triangle loop contributions to the $X(3872)\to\pi^0\chi_{c1,2}$, the ratio of $\Gamma(X(3872)\to\pi^0\chi_{c2})$ to $\Gamma(X(3872)\to\pi^0\chi_{c1})$ was reported to be $1.09$--$1.43$ \cite{wu2021EPJC81-193}. We plot in Fig. \ref{fig:threepoint2ratio} the ratio $\Gamma(X_2\to\pi^0\chi_{c2})/\Gamma(X_2\to\pi^0\chi_{c1})$ with varying the phase angle. Suppose the processes $X_2\to\pi^0\chi_{c1,2}$ by the triangle loops in Figs. \ref{fig:feyndiags}(b) and (c) also have such similar relation and then the phase angle is limited to be between $29.6^\circ$ and $32.1^\circ$ (see Fig. \ref{fig:threepoint2ratio}). In this range, the widths for the $X_2\to\pi^0\chi_{c1,2}$ are\footnote{The superscript symbols $\triangle,\,\bigcirc$, and $\square$ indicate the triangle, bubble, and box loop contributions, respectively.}
\begin{subequations}
	\begin{align}
		\Gamma^{\triangle}(X_2\to\pi^0\chi_{c1})=2.8\sim 21.3 ~\mathrm{keV}\,,\\
		\Gamma^{\triangle}(X_2\to\pi^0\chi_{c2})=4.0\sim 30.4~\mathrm{keV}\,,
	\end{align}
\end{subequations}
corresponding a fraction of $10^{-3}\sim 10^{-2}$ in view of the measured total width of the $X_2$ candidate, $\Gamma_\mathrm{t}(X_2)=4~\mathrm{MeV} $ \cite{wang2022PRD105-112011}. 
In the case of the $X(3872)$, the Particle Data Group \cite{workman2022PTEP2022-083C01} gives $\Gamma(X(3872)\to\pi^0\chi_{c1}) = (40.5\pm 19.0)~\mathrm{keV}$ and a upper limit of $\Gamma (X(3872)\to\pi^0\chi_{c2})<47.6~\mathrm{keV}$.

\begin{figure}
	\centering
	\includegraphics[width=0.92\linewidth]{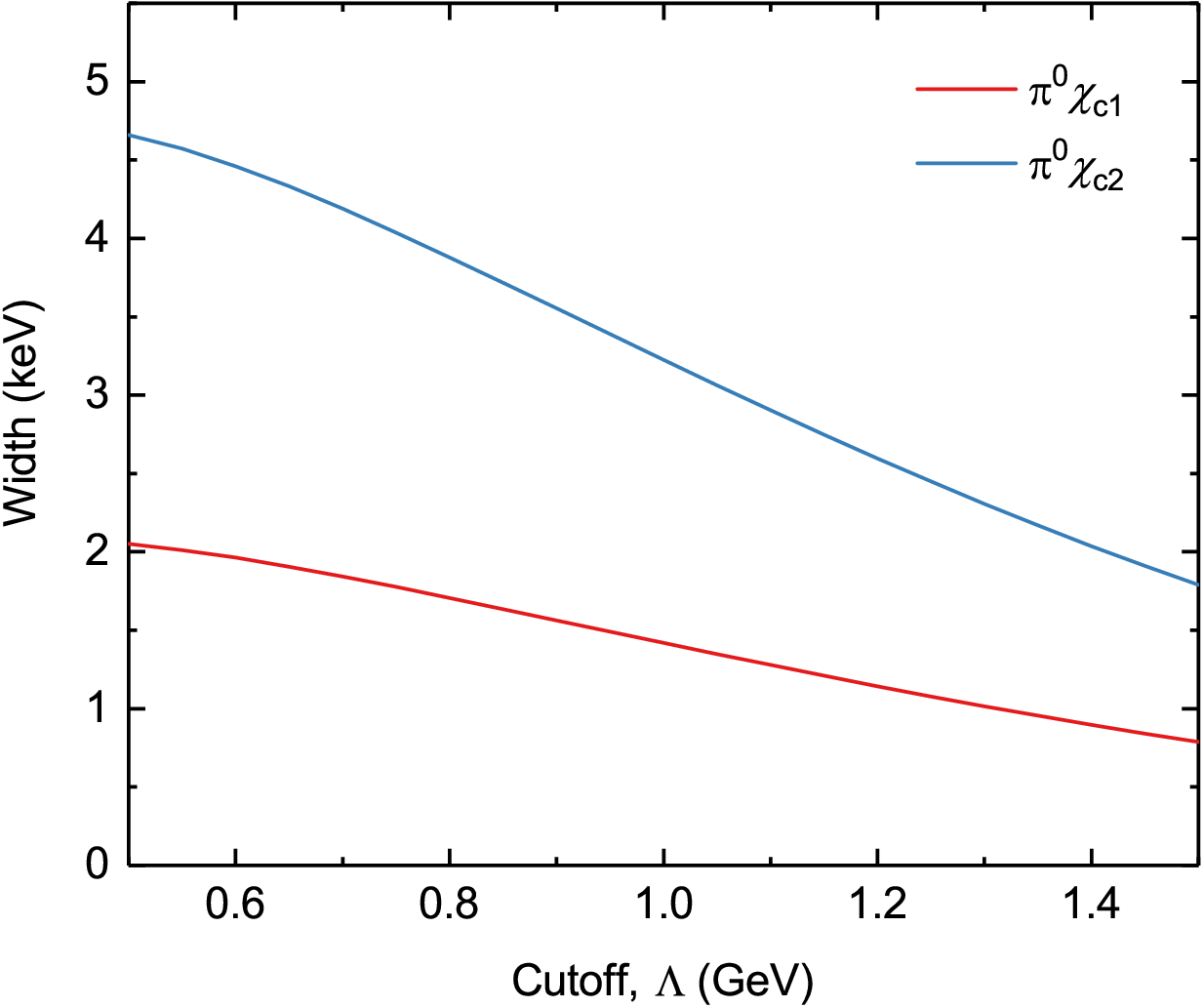}
	\caption{Partial widths for the processes $X_2\to\pi^0\chi_{c1,2}$ via the two-point bubble loop in Fig. \ref{fig:feyndiags}(a). The calculations were performed using $\theta = 31^\circ$ and $c_1=g_1/m_{D^*}$.}
	\label{fig:tploopres}
\end{figure}

Next, we return to the two-point bubble loop in Fig.~\ref{fig:feyndiags}(a), which also contributes to the isospin violated decays $X_2\to\pi^0\chi_{c1,2}$. In the calculations, the cutoff $\Lambda$ is varied from 0.5 to 1.5 GeV \cite{jia2024PRD109-034017} and the coupling constant $c_1$ is fixed to be $g_1/m_{D^*} $, i.e., about $-0.74~\mathrm{GeV^{-3/2}}$. Moreover, the phase angle $\theta$ is chosen to be $31^\circ$. The calculated widths for different cutoffs are presented in Fig. \ref{fig:tploopres}, which are
\begin{subequations}
 \begin{align}
\Gamma^{\bigcirc}(X_2\to\pi^0\chi_{c1})&=0.8\sim 2.1 ~\mathrm{keV}, \\
\Gamma^{\bigcirc}(X_2\to\pi^0\chi_{c2})&=1.8\sim 4.6 ~\mathrm{keV}.
\end{align}    
\end{subequations}
It is seen that with the parameters above, the contribution from the bubble loop is approximately of the same order with that from the triangle loops. This indicates that if the $(c_1 m_{D^*}/g_1)$ is of order unity, the bubble loop contribution to the isospin violated decays $X_2\to\pi^0\chi_{c1,2}$ is somewhat comparable to the triangle loop and should not be neglected in realistic analysis. Clearly, when ($c_1 m_{D^*}/g_1$) is much greater than unity, the contribution from bubble loop would be enhanced. A short discussion on the parameter $c_1$ can be found in Appendix \ref{app:pc}.

\subsection{Partial decay widths of $X_2\to\pi\pi\chi_{c1,2}$}

We firstly consider the partial widths of the processes $X_2\to\pi\pi\chi_{c1,2}$ via the triangle loops in Figs. \ref{fig:feyndiags}(d)--(f). As above, in the calculations we fixed the phase angle $\theta=31^\circ$ and the coupling constant $c_1=-0.74~\mathrm{GeV^{-3/2}}$. Under these conditions, the width for the $X_2\to\pi^0\pi^0\chi_{c1}$ due to the triangle loops is 
\begin{equation}
	\Gamma^{\triangle}(X_2\to\pi^0\pi^0\chi_{c1})=0.36~\mathrm{keV}\,,
\end{equation}
and for the $X_2\to\pi^0\pi^0\chi_{c2}$ it is 
\begin{equation}
	\Gamma^{\triangle}(X_2\to\pi^0\pi^0\chi_{c2})=0.40~\mathrm{keV}\,.
\end{equation}

For the cases of the charged pion emission, the widths via triangle loops are
\begin{subequations}
	\begin{align}
		\Gamma^{\triangle}(X_2\to\pi^+\pi^-\chi_{c1})&=0.63~\mathrm{keV}\,,\\
		\Gamma^{\triangle}(X_2\to\pi^+\pi^-\chi_{c2})&=0.69~\mathrm{keV}\,.
	\end{align}
\end{subequations}

The processes $X_2\to\pi\pi\chi_{c1,2}$ can also occur via the box diagrams shown in Figs. \ref{fig:feyndiags}(g)--(k). The widths for the neutral dipion emission due to the box loops are 
\begin{subequations}
	\begin{align}
		\Gamma^{\square}(X_2\to\pi^0\pi^0\chi_{c1})&=5.2~\mathrm{keV}\,,\\
		\Gamma^{\square}(X_2\to\pi^0\pi^0\chi_{c2})&=4.3~\mathrm{keV}\,.
	\end{align}
\end{subequations}
For the charged dipion emission, the widths are
\begin{subequations}
	\begin{align}
		\Gamma^{\square}(X_2\to\pi^+\pi^-\chi_{c1})&=8.4~\mathrm{keV}\,,\\
		\Gamma^{\square}(X_2\to\pi^+\pi^-\chi_{c2})&=7.1~\mathrm{keV}\,.
	\end{align}
\end{subequations}
We can see that $\Gamma^{\square}(X_2\to\pi^+\pi^-\chi_{c1,2})/\Gamma^{\square}(X_2\to\pi^0\pi^0\chi_{c1,2})\approx 1.6$, in accordance with the expected value of $2$. Moreover, we see that the dipion processes $X_2\to\pi\pi\chi_{cJ}$ occur dominantly via the box loops if the $(c_1 m_{D^*}/g_1)\sim 1$. The relative size of the contributions from the loops in Fig. \ref{fig:feyndiags} could be estimated by the power counting rule (see Appendix \ref{app:pc}), whose results support the numerical calculations above. For easy reference, we summarize in Table \ref{tab:widths} the partial decay widths of the single-pion and dipion processes $X_2\to\pi^0\chi_{cJ}$ and $X_2\to\pi\pi\chi_{cJ}$ due to the diagrams in Fig. \ref{fig:feyndiags}.

\begin{table}
	\caption{Partial decay widths for the processes $X_2\to\pi^0\chi_{cJ} $ and $X_2\to\pi\pi\chi_{cJ}$ due to different loops in Fig. \ref{fig:feyndiags}. The calculations were obtained using $m_{X_2} = 4014.3~\mathrm{MeV}$, $\theta=31^\circ$, $c_1=g_1/m_{D^*}$, and $\Lambda = 1.0~\mathrm{GeV}$. The widths are in units of keV.}
	\label{tab:widths}
	\begin{ruledtabular}
		\begin{tabular}{lccc}
			Final state	&Bubble& Triangle&Box\\
			\colrule
			$\pi^0\chi_{c1}$ & $1.4$ & $6.6$ & $\cdots$\\
			$\pi^0\chi_{c2}$ & $3.2$ & $7.7$ & $\cdots$\\
			$\pi^0\pi^0\chi_{c1}$ & $\cdots$ & $0.36$ & $5.2$\\
			$\pi^0\pi^0\chi_{c2}$ & $\cdots$ & $0.40$ & $4.3$\\
			$\pi^+\pi^-\chi_{c1}$ & $\cdots$ & $0.63$ & $8.4$\\
			$\pi^+\pi^-\chi_{c2}$ & $\cdots$ & $0.69$ & $7.1$\\
		\end{tabular}
	\end{ruledtabular}
\end{table}

\begin{figure}
	\centering
	\includegraphics[width=0.92\linewidth]{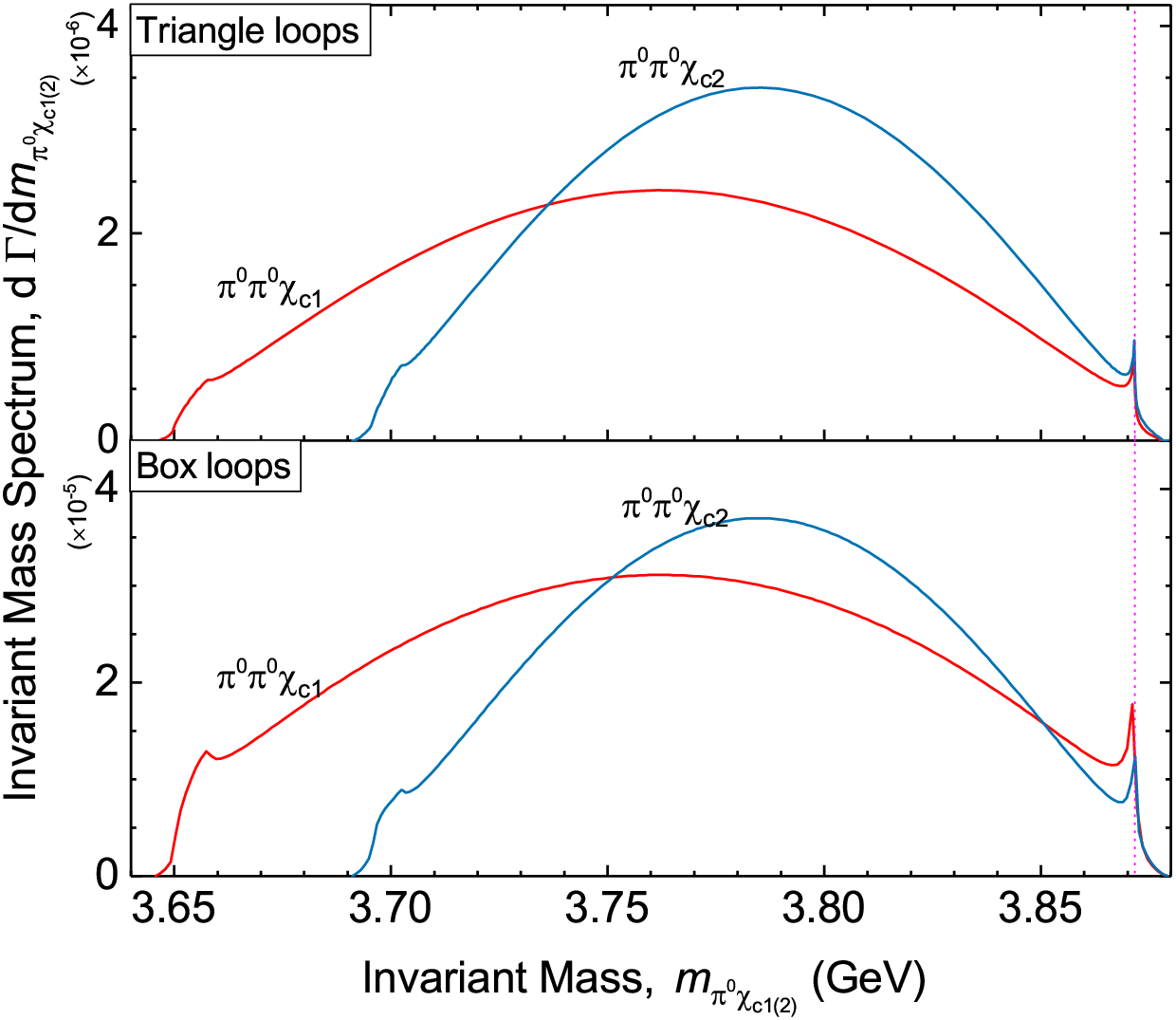}
	\caption{Invariant mass spectrum of the $\chi_{cJ}$ and $\pi^0$ in the processes $X_2\to\pi^0\pi^0\chi_{cJ}$. As indicated, the upper panel presents the contributions from the triangle loops, while the lower panel exhibits those due to the box ones. The vertical dotted line indicates the $D^{*}\bar{D}$ threshold. The results were obtained using $\theta=31^\circ$ and $c_1=-0.74~\mathrm{GeV^{-3/2}}$.}
	\label{fig:dgdm}
\end{figure}

To provide more helpful information for constructing the $X_2$ in experiments, we calculated the $\pi^0\chi_{cJ}$ invariant mass distribution in the dipion processes $X_2\to\pi^0\pi^0\chi_{cJ}$, as shown in Fig.~\ref{fig:dgdm}. A small narrow peak just staying at the $D {\bar D}^*$ threshold--3872 MeV can be clearly seen in Fig.~\ref{fig:dgdm}. This is the cusp structure enhanced and narrowed by the nearby triangle singularity (TS). 
It implies that even without introducing a genuine resonance, the cusp enhanced by the nearby TS can still simulate a narrow resonance-like structure. The threshold cusp could be used to describe some resonance-like structures observed in experiments \cite{liu2019PRD100-054006,chen2011PRD84-094003,bugg2011E96-11002,xie2019PLB792-450}. Consequently, this cusp structure might hint that within the molecular framework, the $X_2$, as the spin-2 $D^{*}\bar{D}^{*}$ of the molecule state $X(3872)$, probably first decays into $D^{*}$ and $\bar{D}^{*}$, then the $D^{*}$ ($\bar{D}^{*}$) goes to $D$ ($\bar{D}$) with emission of a pion, and finally the produced $D$ ($\bar{D}$) and foregoing $\bar{D}^{*}$ ($D^{*}$) could probably combine together to form the $X(3872)$, which would further decay with some probability into the $\chi_{c1,2}$ and one pion. Maybe, we could reconstruct the $X_2$ using some possible processes where the $X(3872)$ plus one neutral pion were observed.

\subsection{$X_2$ mass influence}

In terms of Eq. \eqref{eq:coupgx2}, the coupling strength of the $X_2$ to its components depends strongly on the $X_2$ mass. Hence, in the following we focus on the influence of the $X_2$ mass on the processes $X_2\to\pi\pi\chi_{cJ}$. In view of the mass of the resonance measured by the Belle collaboration: $m_{X_2}=(4014.3\pm 4.0\pm 1.5)~\mathrm{MeV}$ \cite{wang2022PRD105-112011}, we vary the $X_2$ mass from $4008.8~\mathrm{MeV}$ to $4019.8~\mathrm{MeV}$ in our calculations. We consider only the box diagrams to show the influence of the $X_2$ mass on the dipion processes.

\begin{figure}
	\centering
	\includegraphics[width=0.92\linewidth]{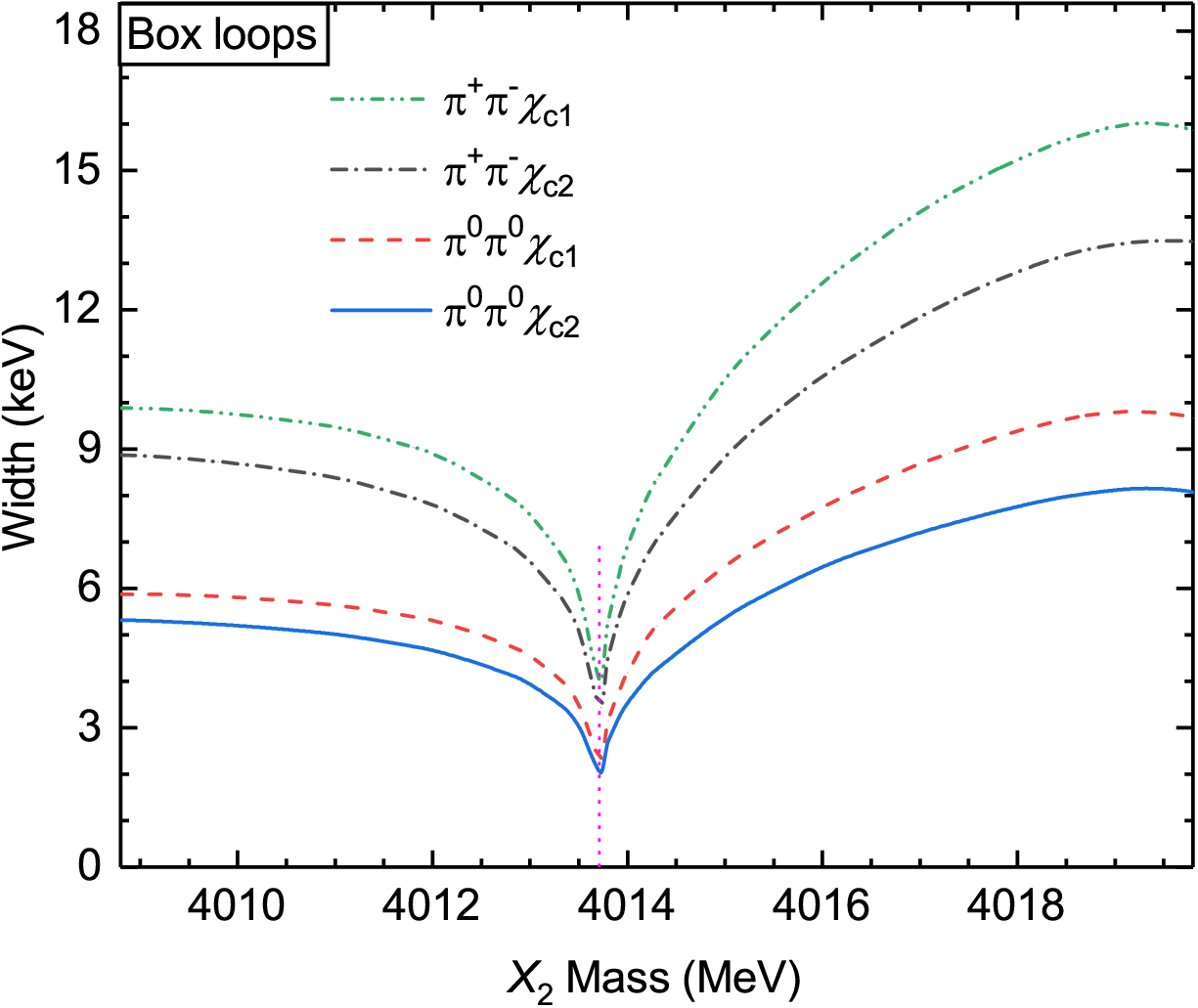}
	\caption{Dependence of the partial widths for the $X_2\to\pi\pi\chi_{cJ}$ on the mass of the $X_2$. The vertical dotted line indicates the $D^{*0}\bar{D}^{*0}$ threshold. In the calculations, the mixing angle $\theta=31^\circ$ and only the box diagrams were considered.}
	\label{fig:widthdipionvsmx2}
\end{figure}

\begin{figure}
	\centering
	\includegraphics[width=0.92\linewidth]{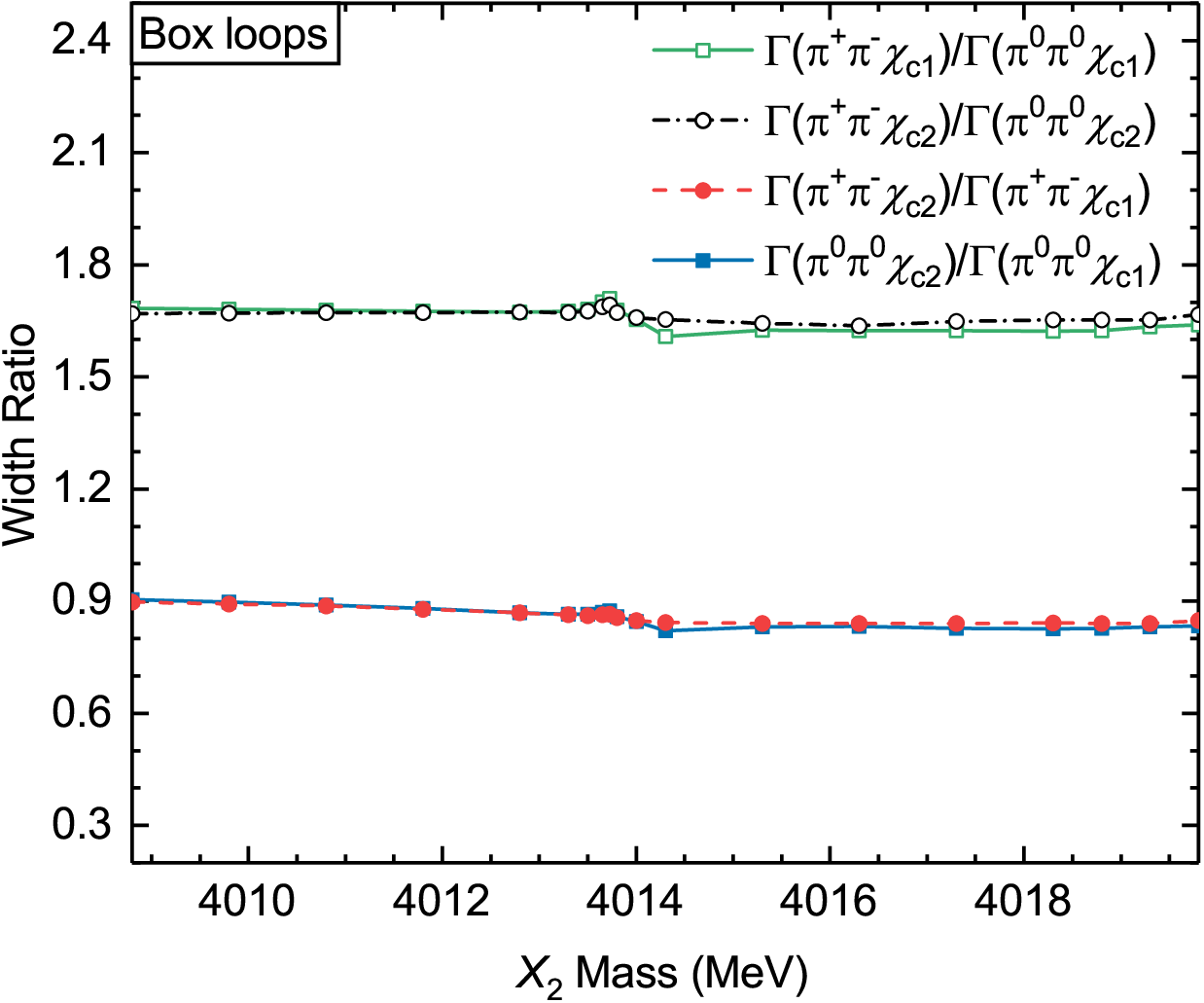}
	\caption{Width ratios for the dipion processes $X_2\to\pi\pi\chi_{cJ}$ with varying the $X_2$ mass. Again, the mixing angle $\theta=31^\circ$ and only the box diagrams were considered in the calculations.}
	\label{fig:widthratiodipion}
\end{figure}

In Fig. \ref{fig:widthdipionvsmx2} the partial decay widths for the $X_2\to\pi\pi\chi_{cJ}$ are plotted as a function of the $X_2$ mass. It is
seen that the partial decay widths for the different processes we considered here show a similar behavior. At the $D^{*0}\bar{D}^{*0}$ threshold, i.e., $m_{X_2}=4013.7~\mathrm{MeV}$, the curves show a valley. This is straightforward since at $m_{X_2}=4013.7~\mathrm{MeV}$ the coupling of the $X_2$ to the $D^{*0}\bar{D}^{*0}$ is zero according to Eq. \eqref{eq:coupgx2}. Although the partial widths are sensitive to the $X_2$ mass, the ratios between the widths for the $X_2\to\pi\pi\chi_{cJ}$, defined as
\begin{subequations}\label{eq:widthratio}
	\begin{align}
		R_{21} &= \frac{\Gamma (X_2\to\pi\pi\chi_{c2})}{\Gamma (X_2\to\pi\pi\chi_{c1})},\,\\
		R_{\mathrm{c}0} &= \frac{\Gamma (X_2\to\pi^+\pi^-\chi_{cJ})}{\Gamma (X_2\to\pi^0\pi^0\chi_{cJ})},\,
	\end{align}
\end{subequations}
are nearly independent of the $X_2$ mass (see Fig. \ref{fig:widthratiodipion}). It shows that
\begin{subequations}
	\begin{align}
		R_{21} &\approx 0.85,\,\\
		R_{\mathrm{c}0} &\approx 1.6,\,
	\end{align}
\end{subequations}

For the isospin broken processes $X_2\to\pi^0\chi_{cJ}$, the decay widths are quite sensitive to the $X_2$ mass as well as the mixing angle. This is because the effective coupling strength of the $X_2$ to its components $D^*\bar{D}^*$ depends on the mixing angle and the $X_2$ mass [see Eqs. \eqref{eq:Lx2} and \eqref{eq:coupgx2}]. It is recalled that in the foregoing section we determined the mixing angle $\theta=29.6^\circ\sim32.1^\circ$ using the $X_2$ mass $m_{X_2}=4014.3~\mathrm{MeV}$. Under these parameters, the effective coupling strength
\begin{equation}\label{eq:constraint}
	 \frac{g_X^\mathrm{n}\cos\theta}{g_X^\mathrm{c}\sin\theta} = 0.89\sim 0.98\,.
\end{equation}
This relation is tantamount to assuming that the breaking of the isospin for the processes $X_2\to\pi^0\chi_{cJ}$ results dominantly from the mass difference between the neutral and charged charmed mesons (or the $u$ and $d$ quarks). 

\begin{figure}[t]
	\centering
	\includegraphics[width=0.92\linewidth]{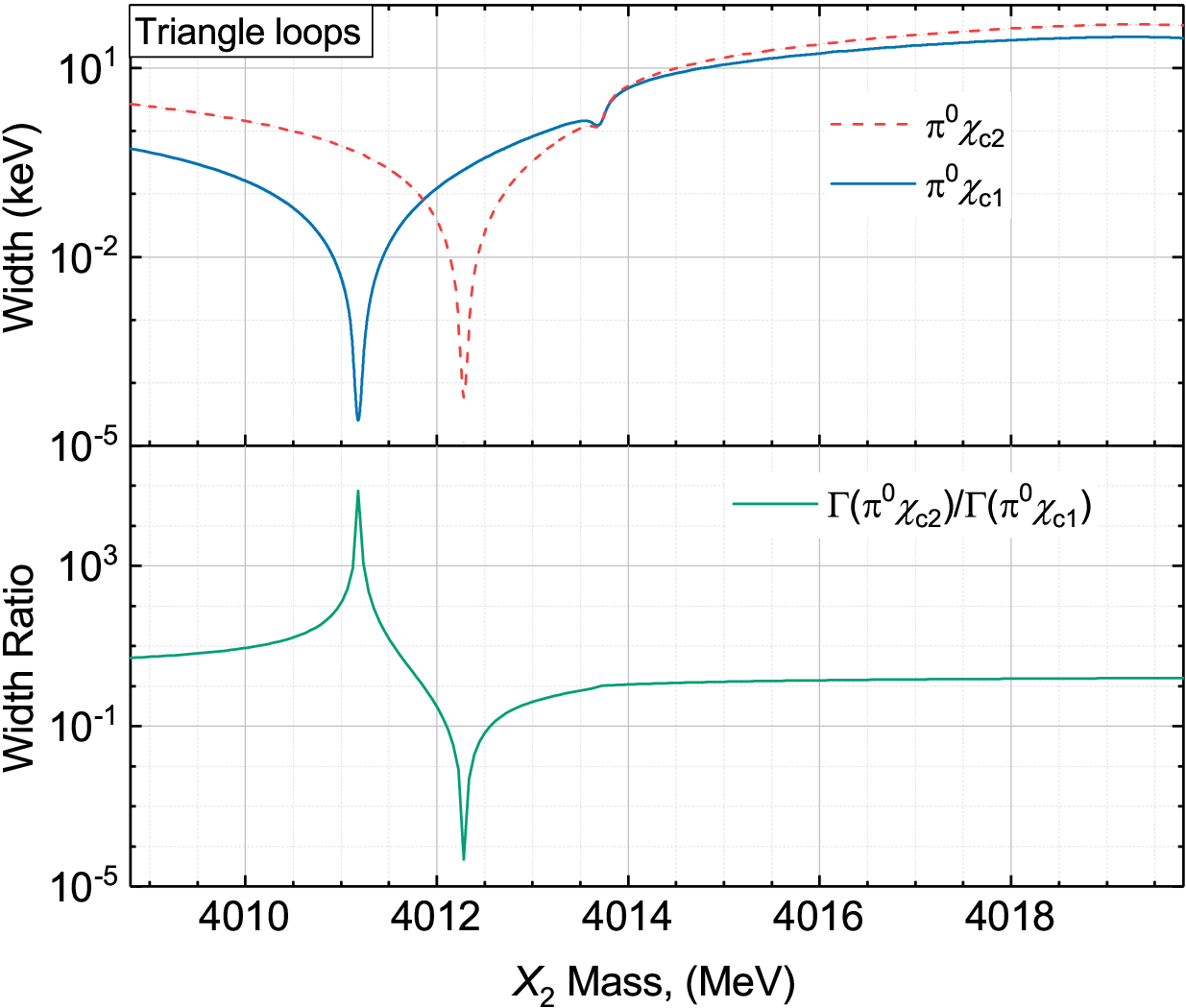}
	\caption{Partial widths of the $X_2\to\pi^0\chi_{cJ}$ (upper) and the corresponding width ratio (lower) for different mass of the $X_2$. In the calculations, we fixed $(g_X^\mathrm{n}\cos\theta)/(g_X^\mathrm{c}\sin\theta )\approx0.93$ and considered only the triangle loops.}
	\label{fig:widthratiosinglepion}
\end{figure}

If we maintain the constraint described by Eq. \eqref{eq:constraint}, the partial widths for the single pion processes $X_2\to\pi^0\chi_{cJ}$ for different $X_2$ mass are plotted in Fig. \ref{fig:widthratiosinglepion}, exhibiting different behavior from the dipion processes $X_2\to\pi\pi\chi_{cJ}$.

\subsection{Comparison with $X(3872)$}

In absence of the experimental data for the $X_2\to\pi(\pi)\chi_{cJ}$, we cannot make a direct comparison with our model calculations. However, some comparisons between the cases of the $X(3872)$ and $X_2$ are available because there have been many experimental measurements and theoretical calculations for the $X(3872)$. In Table \ref{tab:widthratio}, we list the available ratios related to the $X(3872)\to\pi(\pi)\chi_{cJ}$, together with those for the $X_2$ we calculated in this work.

From Table \ref{tab:widthratio}, we see that under the molecular picture, the isospin-conserved dipion processes for the $X(3872)$ are much weaker than the isospin-broken single-pion processes, whereas for the $X_2$ the dipion and single-pion decays are of approximately equal rate. Another difference is that the charged and the neutral dipion transitions of the $X_2$ are in line with the expected behavior, namely $\Gamma (X_2\to\pi^+\pi^-\chi_{cJ})/\Gamma (X_2\to\pi^0\pi^0\chi_{cJ})\approx 2 $, whereas the $X(3872)$ exhibits the nearly same decay rate when assuming the $X(3872)$ as a conventional charmonium state $\chi_{c1}(2{}^3P_1)$ \cite{achasov2024x-}. Additionally, the rates of the $X(3872)$ 
or $X_2$ decaying into $\pi^0\chi_{c1}$ and $\pi^0\chi_{c2}$ are of the same order. Here, it should be pointed out that according to Refs. \cite{achasov2024PRD109-036028,achasov2024x-} the large ratio of 25 for the $\Gamma(X(3872)\to\pi^+\pi^-\chi_{c1})/\Gamma (X(3872)\to\pi^0\chi_{c1})$ can be attributed to the insignificant two-gluon production mechanism of the $\pi^0$, in comparison with the intermediate charmed meson loops. Actually, when taking into account the charmed meson loops, the isospin-broken single pion decay $X(3872)\to\pi^0\chi_{c1}$ would be strongly enhanced \cite{achasov2024PRD109-036028}, leading to a comparable value for the $\Gamma(X(3872)\to\pi^+\pi^-\chi_{c1})/\Gamma (X(3872)\to\pi^0\chi_{c1})$ to that obtained within the molecule model (see Table \ref{tab:widthratio}).

\begin{table}
	\caption{Width ratios related to the processes $X(3872)\to\pi(\pi)\chi_{cJ}$ and $X_2\to\pi(\pi)\chi_{cJ}$. The data for the $X_2$ was obtained at $m_{X_2}=4014.3~\mathrm{MeV}$. For the dipion decays, we consider only the box loops, and for the single-pion decays, only the triangle loops are taken.}
	\label{tab:widthratio}
	\begin{ruledtabular}
		\begin{tabular}{lllc}
			Ratio	&$X(3872)$& Reference & $X_2$\\
			\colrule
			$(\pi^+\pi^-\chi_{c1})/(\pi^0\chi_{c1})$ & $<0.2$\footnotemark[1]& \cite{ablikim2019PRL122-202001,ablikim2024PRD109-L071101,workman2022PTEP2022-083C01} & $1.3$ \\
			 & $25$\footnotemark[2] &\cite{dubynskiy2008PRD77-014013}&  \\
			 & $(2.5\sim 5.2)\times 10^{-3}$\footnotemark[2]  &\cite{achasov2024x-,achasov2024PRD109-036028}&  \\
			 & $\order{10^{-3}}$\footnotemark[3]  &\cite{fleming2008PRD78-094019}& \\
			 $(\pi^0\pi^0\chi_{c1})/(\pi^0\chi_{c1})$& $0.61 $\footnotemark[3] &\cite{fleming2008PRD78-094019}&$0.79$\\
			 & $6.7\times 10^{-2} $\footnotemark[3] &\cite{dong2009PRD79-094013}&\\
			 & $2.9\times 10^{-3} $\footnotemark[3] &\cite{fleming2012PRD85-014016}&\\
			 $(\pi^0\pi^0\chi_{c2})/(\pi^0\chi_{c2})$&$7.8\times 10^{-6}$\footnotemark[3]&\cite{fleming2008PRD78-094019}&0.56\\
			 &$1.4\times 10^{-3}$\footnotemark[3]&\cite{dong2009PRD79-094013}&\\
             $(\pi\pi\chi_{c2})/(\pi\pi\chi_{c1})$&$10^{-4}$\footnotemark[2]& \cite{dubynskiy2008PRD77-014013}&$0.85$ \\
             & $\sim 10^{-6}$\footnotemark[3]& \cite{fleming2008PRD78-094019}& \\
             $(\pi^+\pi^-\chi_{c1})/(\pi^0\pi^0\chi_{c1})$& $1.1$\footnotemark[2] &\cite{achasov2024x-} & $1.6$\\
             $(\pi^0\chi_{c2})/(\pi^0\chi_{c1})$ & $<1.2$\footnotemark[1] &\cite{workman2022PTEP2022-083C01} & $1.2$\\
             & $1.1$--$1.4$\footnotemark[3]  &\cite{wu2021EPJC81-193}& \\
             & $0.37$\footnotemark[2]  &\cite{dubynskiy2008PRD77-014013}& \\
             & $0.64$\footnotemark[3] &\cite{fleming2008PRD78-094019}& \\
             & $0.83$\footnotemark[3] &\cite{mehen2015PRD92-034019}& \\
             & $0.77$\footnotemark[3] &\cite{zhou2019PRD100-094025}& \\
             & $\sim 1.0$\footnotemark[4] &\cite{dubynskiy2008PRD77-014013}& \\
		\end{tabular}
	\end{ruledtabular}
	\footnotetext[1]{The experimental measurements.}
	\footnotetext[2]{Theoretical predictions under the $X(3872)$ as the conventional charmonium state $\chi_{c1}(2{}^3P_1)$.}
	\footnotetext[3]{Theoretical predictions under the $X(3872)$ as the $D^*\bar{D}$ molecule.}
	\footnotetext[4]{Theoretical predictions under the $X(3872)$ as the tetraquark state.}
\end{table}

\section{Summary}
\label{sec:summary}
We calculated the single-pion and dipion transitions of the $X_2$ to the $\chi_{c1,2}$ using a nonrelativistic effective field theory. In the calculations, we treated the $X_2$ as a $D^{*}\bar{D}^*$ molecule and assumed the pionic processes occur through the intermediate charmed meson loops. In analogy with the $X(3872)\to\pi^0\chi_{cJ}$, the phase angle that describes the proportion of the neutral and charged $D^{*}\bar{D}^{*}$ in the $X_2$ is determined to be about $31^\circ$.

Taking a reasonable coupling constant for the $D^*\bar{D}^*\chi\pi$ vertex, the single-pion processes $X_2\to\pi^0\chi_{cJ}$ are dominantly governed by the triangle loops, whereas the dipion processes $X_2\to\pi\pi\chi_{cJ}$ results mainly from the box diagrams. Our model shows that the widths for the single-pion processes $X_2\to\pi^0\chi_{cJ}$ are about 3--30 keV, and for the dipion decays $X_2\to\pi\pi\chi_{cJ}$ they are between 4 keV and 9 keV. Taking the total width $\Gamma_\mathrm{t} =4~\mathrm{MeV}$ for the $X_2$, the corresponding branching fractions are between $10^{-3}$ and $10^{-2}$. The width ratios $R_{\mathrm{c}0}=\Gamma (X_2\to\pi^+\pi^-\chi_{cJ})/\Gamma (X_2\to\pi^0\pi^0\chi_{cJ}) \simeq 1.6$, somewhat smaller than the expected value of $2$, and $R_{21}=\Gamma (X_2\to\pi\pi\chi_{c2})/\Gamma (X_2\to\pi\pi\chi_{c1}) \simeq 0.85$. The two ratios remain nearly unchanged when varying the $X_2$ mass. We also presented the invariant mass spectra of the $\pi^0\chi_{cJ}$ final state in the $X_2\to\pi^0\pi^0\chi_{cJ}$, showing a small narrow cusp at the $D^{*}\bar{D}$ threshold enhanced and narrowed by the nearby triangle singularity. The spectral shape can be used to compare with the future experimental results. To some extent, the threshold cusp effect gives us a hint that we could reconstruct the $X_2$ using some possible processes where the $X(3872)$ plus one neutral pion were observed. We also give a short comparison between the $X(3872)$ and $X_2$.

\begin{acknowledgements}\label{sec:acknowledgements}
The author (S.D. Liu) are grateful to Zhen-Hua Zhang for constructive instruction and discussion. This work is partly supported by the National Natural Science Foundation of China under Grant Nos. 12105153, 12475081, 12075133, 12075288, 12235018 and 11975165, and by the Natural Science Foundation of Shandong Province under Grant Nos. ZR2021MA082 and ZR2022ZD26. It is also supported by Taishan Scholar Project of Shandong Province (Grant No.tsqn202103062), the Higher Educational Youth Innovation Science and Technology Program Shandong Province (Grant No. 2020KJJ004).	
\end{acknowledgements}

\onecolumngrid
\appendix
\section{Loop Integrals}\label{app:loopfunc}
The loop integral functions we used are the same with those in Refs. \cite{chen2020CPC44-023103,chen2017PRD95-034022,jia2024PRD109-034017,guo2011PRD83-034013} so that we only give the relevant formulas.
In the rest frame of the $X_2$ [$p=(M,\,\vec{0})$],
the scalar integral for the bubble loops is \cite{guo2011PRD83-034013,jia2024PRD109-034017}
\begin{equation}\label{eq:apptwoloop}
	\begin{split}
		I(m_1,\,m_2) &\equiv \ii \int \frac{\dd[4]{l}}{(2\pi)^4} \frac{1}{(l^2-m_1^2+\ii \epsilon)[(p-l)^2-m_2^2+\ii \epsilon]}\\
		& = \frac{\mu_{12}}{4\pi^2 m_1m_2} \Big[\Lambda - \sqrt{c_{12}-\ii\epsilon}\arctan\Big(\frac{\Lambda}{\sqrt{c_{12}-\ii\epsilon}}\Big)\Big]\,,
	\end{split}
\end{equation}
where $\mu_{12}$ is the reduced mass of the intermediate mesons $D^*$ and $\bar{D}^{*}$, and the parameter 
\begin{equation}
	c_{12} = 2\mu_{12}(m_1+m_2-M)\,.
\end{equation}

The scalar integral for triangle loops reads \cite{guo2011PRD83-034013,jia2024PRD109-034017}
\begin{equation}
	\begin{split}
		I(m_1,\,m_2,\,m_3)&\equiv\ii\int \frac{\dd[4]{l}}{(2\pi)^4} \frac{1}{(l^2-m_1^2+\ii \epsilon)[(p-l)^2-m_2^2+\ii \epsilon][(l-q)^2-m_3^2+\ii\epsilon]}\\
		&=\frac{\mu_{12}\mu_{23}}{16\pi m_1m_2m_3} \frac{1}{\sqrt{a}} \Big[\arctan\Big(\frac{c_{23}-c_{12}}{2\sqrt{ac_{12}}}\Big)+\arctan\Big(\frac{2a+c_{12}+c_{23}}{2\sqrt{a(c_{c23}-a)}}\Big)\Big]\,.
	\end{split}
\end{equation}
Here
\begin{subequations}
	\begin{align}
		a&\equiv\left(\frac{\mu_{23}}{m_3}\right)^2\bm{q}^2\,,\\
		c_{23}&\equiv 2\mu_{23}(m_2+m_3+q^0-M)+\frac{\mu_{23}}{m_3} \bm{q}^2
	\end{align}
\end{subequations}

\begin{figure}[htbp]
	\centering
	\includegraphics[width=0.56\linewidth]{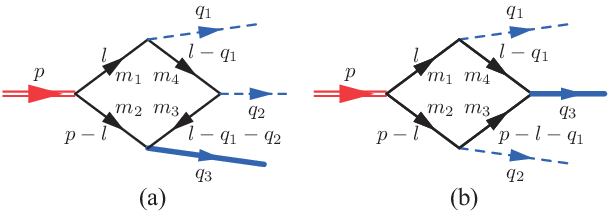}
	\caption{The momentum labels for the box loops considered in present work.}
	\label{fig:flpm}
\end{figure}

For the box loop in Fig. \ref{fig:flpm}(a), the integral is written as \cite{chen2020CPC44-023103,chen2017PRD95-034022,jia2024PRD109-034017}
\begin{equation}\label{eq:Im14a}
	\begin{split}
		I(m_i) &\equiv \ii\int \frac{\dd[4]{l}}{(2\pi)^4} \frac{1}{(l^2-m_1^2+\ii\epsilon)[(p-l)^2-m_2 +\ii\epsilon][(l-q_1-q_2)^2-m_3^2+\ii\epsilon][(l-q_1)^2-m_4^2+\ii\epsilon]}\\
		& = - \frac{\mu_{12}\mu_{23}\mu_{24}}{2m_1m_2m_3m_4}\int \frac{\dd[3]{\bm{l}}}{(2\pi)^3}\frac{1}{(\vec{l}^2+c_{12}-\ii\epsilon) (\bm{l}^2+2\frac{\mu_{23}}{m_3}\bm{l}\cdot\bm{q}_3 +c_{23}'-\ii\epsilon)(\bm{l}^2-2\frac{\mu_{24}}{m_4}\bm{l}\cdot\bm{q}_1+c_{24}-\ii\epsilon)}\,.
	\end{split}
\end{equation}
Here and later the $I(m_i)$ stands for $I(m_1,\,m_2,\,m_3,\,m_4)$. For this loop, there is a corresponding crossed diagram when the two light outgoing particles are $\pi^0$'s. In this case, the loop can be obtained by swapping $q_1$ and $q_2$ and its integral becomes
\begin{equation}\label{eq:Im14b}
		I(m_i) = - \frac{\mu_{12}\mu_{23}\mu_{24}}{2m_1m_2m_3m_4}\int \frac{\dd[3]{\bm{l}}}{(2\pi)^3}\frac{1}{(\vec{l}^2+c_{12}-\ii\epsilon) (\bm{l}^2+2\frac{\mu_{23}}{m_3}\bm{l}\cdot\bm{q}_3 +c_{23}'-\ii\epsilon)(\bm{l}^2-2\frac{\mu_{24}}{m_4}\bm{l}\cdot\bm{q}_2+c_{24}'-\ii\epsilon)}\,.
\end{equation}

The integral for the loop in Fig. \ref{fig:flpm}(b) is given by \cite{chen2020CPC44-023103,chen2017PRD95-034022,jia2024PRD109-034017}
\begin{equation}\label{eq:Im14c}
	\begin{split}
		I(m_i) &\equiv  \ii \int \frac{\dd[4]{l}}{(2\pi)^4}\frac{1}{(l^2-m_1^2+\ii\epsilon)[(p-l)^2-m_2^2+\ii\epsilon][(p-l-q_2)^2-m_3^2+\ii\epsilon][(l-q_1)^2-m_4^2+\ii\epsilon]}\\
		&=-\frac{\mu_{12}\mu_{34}}{2m_1m_2m_3m_4}\int \dfrac{\dd[3]{\bm{l}}}{(2\pi)^3}\frac{1}{(\bm{l}^2+c_{12}-\ii\epsilon)(\bm{l}^2-\frac{2\mu_{34}}{m_4}\bm{l}\cdot\bm{q}_1+\frac{2\mu_{34}}{m_3}\bm{l}\cdot\bm{q}_2+c_{34}-\ii\epsilon)}\\
		&\times\Big(\frac{\mu_{24}}{\bm{l}^2-\frac{2\mu_{24}}{m_4}\bm{l}\cdot\bm{q}_1+c_{24}-\ii\epsilon}+\frac{1}{\bm{l}^2+\frac{2\mu_{13}}{m_3}\bm{l}\cdot\bm{q}_2+c_{13}-\ii\epsilon}\Big)\,.
	\end{split}
\end{equation}
The parameters $c$'s in Eqs. \eqref{eq:Im14a}--\eqref{eq:Im14b} are defined as
\begin{subequations}
	\begin{align}
		c_{23}'&\equiv 2\mu_{23}(m_2+m3-M+q_1^0+q_2^0)+\frac{\mu_{23}}{m_3}\bm{q}_3^2\\
		c_{24}&\equiv 2\mu_{24}(m_2+m_4-M+q_1^0)+\frac{\mu_{24}}{m_4}\bm{q}_1^2\\
		c_{24}'&\equiv 2\mu_{24}(m_2+m_4-M+q_2^0)+\frac{\mu_{24}}{m_4}\bm{q}_2^2\\
		c_{34} &\equiv 2\mu_{34}(m_3+m_4-q_3^0)+\frac{\mu_{34}}{m_4}\bm{q}_1^2+\frac{\mu_{34}}{m_3}\bm{q}_2^2\\
		c_{13} &\equiv 2\mu_{13}(m_1+m_3-M+q_2^0)+\frac{\mu_{13}}{m_3}\bm{q}_2^2
	\end{align}
\end{subequations}

\section{Power Counting Estimations of Loop Contributions}\label{app:pc}
The power counting estimations here are similar to those in Refs. \cite{guo2011PRD83-034013,guo2010PRD82-034025,jia2024PRD109-034017}. The integral functions in Appendix \ref{app:loopfunc} can be counted in the powers of the velocity of the intermediate mesons,
\begin{equation}
	v= \Big(\frac{2\abs*{m_1+m_2-M_{i(f)}}}{m_1+m_2}\Big)^{1/2},\,
\end{equation}
where $m_1$ and $m_2$ are the masses of the charmed mesons related to the initial meson of mass $M_i$ and the final heavy meson of mass $M_f$. For the $X_2$, $v=0.04$, and for the $\chi_{c1,2}$, $v=0.5$. Hence, the average velocity $v=0.27$.

Within the nonrelativistic framework, the energy scales as $v^2$ and the meson propagator also counts as $v^2$. According to the Lagrangians in Eqs. \eqref{eq:Lagpi}, \eqref{eq:LchiDDpi}, and \eqref{eq:LagDDpipi}, the vertices $D^{(*)}D^{(*)}\pi$, $ D^{(*)}\bar{D}^{(*)}\chi_{cJ}\pi$, and $D^{(*)}D^{(*)}\pi\pi$ could scale as the pion momentum $p_\pi$, and the vertices $X_2D^{(*)}\bar{D}^{(*)}$ and $\chi_{cJ}D^{(*)}\bar{D}^{(*)}$ count as $1$ in terms of the Lagrangians in Eqs. \eqref{eq:Lx2} and \eqref{eq:LagchiDD}.

Therefore, for the single-pion processes $X_2\to\pi^0\chi_{cJ}$ the ratio of the contribution from the bubble loop in Fig. \ref{fig:feyndiags}(a) to that from the triangle loops in Fig. \ref{fig:feyndiags}(b) or (c) reads
\begin{equation}
	\frac{c_1m_{D^{(*)}}}{g_1}v^2\,.
\end{equation}
If $(c_1m_{D^{(*)}})/g_1$ is of order unity, the contribution of the bubble loop is suppressed by a factor of $1/v^2\sim \order{10}$. Similarly, for the dipion processes $X_2\to\pi\pi\chi_{cJ}$, the ratio between the triangle and box loop is also
\begin{equation}
		\frac{c_1m_{D^{(*)}}}{g_1}v^2 \sim \order{0.1}, \quad \text{if }~\frac{c_1m_{D^{(*)}}}{g_1} \sim 1\,.
\end{equation}
Compared to our numerical results, the power counting somewhat overestimates the suppression on the bubble loops for the $X_2\to\pi^0\chi_{cJ}$ and on the triangle loops for the $X_2\to\pi\pi\chi_{cJ}$.

Notice that the above estimation depends on the value of the parameter $(c_1m_{D^*}/g_1)$. Once $ (c_1m_{D^*}/g_1) \gg 1 $, the importance of the box loops for the dipion processes and the triangle loops for the single-pion processes would be weakened. By matching the calculated results for the $X(3872)\to\chi_{cJ}\pi^0$ in Refs. \cite{fleming2008PRD78-094019,dubynskiy2008PRD77-014013}, $(c_1m_{D^*}/g_1)$ tends to be about $10~\mathrm{GeV^{-1}}$. In Ref. \cite{achasov2024x-}, the $g_{D^0\bar{D}^0\pi^0\chi_{c1}}$ was taken to be about $137~\mathrm{GeV}$, corresponding to $\abs{c_1}\approx 5.2~\mathrm{GeV^{-3/2}} $, which is about $7$ times greater than the value ($\abs{c_1}=0.74~\mathrm{GeV^{-3/2}}$) we used. Under these values, all the diagrams in Fig. \ref{fig:feyndiags} would be of equal importance based on the power counting rule. It should be pointed out that the larger value of $c_1$ only enhances the importance of the relevant diagrams, but do not significantly change the numerical results, especially for the width ratios defined in Eq. \eqref{eq:widthratio} and the shape of the invariant mass spectra in Fig. \ref{fig:dgdm}, and hence the conclusion remains nearly unchanged.

\twocolumngrid
\bibliography{particlePhys.bib}

\providecommand{\noopsort}[1]{}
\begin{thebibliography}{77}%
\makeatletter
\providecommand \@ifxundefined [1]{%
 \@ifx{#1\undefined}
}%
\providecommand \@ifnum [1]{%
 \ifnum #1\expandafter \@firstoftwo
 \else \expandafter \@secondoftwo
 \fi
}%
\providecommand \@ifx [1]{%
 \ifx #1\expandafter \@firstoftwo
 \else \expandafter \@secondoftwo
 \fi
}%
\providecommand \natexlab [1]{#1}%
\providecommand \enquote  [1]{``#1''}%
\providecommand \bibnamefont  [1]{#1}%
\providecommand \bibfnamefont [1]{#1}%
\providecommand \citenamefont [1]{#1}%
\providecommand \href@noop [0]{\@secondoftwo}%
\providecommand \href [0]{\begingroup \@sanitize@url \@href}%
\providecommand \@href[1]{\@@startlink{#1}\@@href}%
\providecommand \@@href[1]{\endgroup#1\@@endlink}%
\providecommand \@sanitize@url [0]{\catcode `\\12\catcode `\$12\catcode
  `\&12\catcode `\#12\catcode `\^12\catcode `\_12\catcode `\%12\relax}%
\providecommand \@@startlink[1]{}%
\providecommand \@@endlink[0]{}%
\providecommand \url  [0]{\begingroup\@sanitize@url \@url }%
\providecommand \@url [1]{\endgroup\@href {#1}{\urlprefix }}%
\providecommand \urlprefix  [0]{URL }%
\providecommand \Eprint [0]{\href }%
\providecommand \doibase [0]{https://doi.org/}%
\providecommand \selectlanguage [0]{\@gobble}%
\providecommand \bibinfo  [0]{\@secondoftwo}%
\providecommand \bibfield  [0]{\@secondoftwo}%
\providecommand \translation [1]{[#1]}%
\providecommand \BibitemOpen [0]{}%
\providecommand \bibitemStop [0]{}%
\providecommand \bibitemNoStop [0]{.\EOS\space}%
\providecommand \EOS [0]{\spacefactor3000\relax}%
\providecommand \BibitemShut  [1]{\csname bibitem#1\endcsname}%
\let\auto@bib@innerbib\@empty
\bibitem [{\citenamefont {Choi}\ \emph {et~al.}(2003)\citenamefont {Choi},
  \citenamefont {Olsen}, \citenamefont {Abe} \emph
  {et~al.}}]{choi2003PRL91-262001}%
  \BibitemOpen
  \bibfield  {author} {\bibinfo {author} {\bibfnamefont {S.}~\bibnamefont
  {Choi}}, \bibinfo {author} {\bibfnamefont {S.}~\bibnamefont {Olsen}},
  \bibinfo {author} {\bibfnamefont {K.}~\bibnamefont {Abe}}, \emph {et~al.}
  (\bibinfo {collaboration} {Belle}),\ }\bibfield  {title} {\bibinfo {title}
  {Observation of a narrow charmoniumlike state in exclusive
  ${B}^{\ifmmode\pm\else\textpm\fi{}}\ensuremath{\rightarrow}{K}^{\ifmmode\pm\else\textpm\fi{}}{\ensuremath{\pi}}^{+}{\ensuremath{\pi}}^{\ensuremath{-}}{J}/\ensuremath{\psi}$
  decays},\ }\href {https://doi.org/10.1103/PhysRevLett.91.262001} {\bibfield
  {journal} {\bibinfo  {journal} {Phys. Rev. Lett.}\ }\textbf {\bibinfo
  {volume} {91}},\ \bibinfo {pages} {262001} (\bibinfo {year} {2003})},\
  \Eprint {https://arxiv.org/abs/hep-ex/0309032} {arxiv:hep-ex/0309032}
  \BibitemShut {NoStop}%
\bibitem [{\citenamefont {Ablikim}\ \emph {et~al.}(2013)\citenamefont {Ablikim}
  \emph {et~al.}}]{ablikim2013PRL110-252001}%
  \BibitemOpen
  \bibfield  {author} {\bibinfo {author} {\bibfnamefont {M.}~\bibnamefont
  {Ablikim}} \emph {et~al.} (\bibinfo {collaboration} {BESIII}),\ }\bibfield
  {title} {\bibinfo {title} {Observation of a charged charmoniumlike structure
  in
  ${e}^{\mathbf{+}}{e}^{\mathbf{\ensuremath{-}}}\ensuremath{\rightarrow}{\ensuremath{\pi}}^{\mathbf{+}}{\ensuremath{\pi}}^{\mathbf{\ensuremath{-}}}{J}/\ensuremath{\psi}$
  at $\sqrt{s} = 4.26\text{ }\text{ }\mathrm{GeV}$},\ }\href
  {https://doi.org/10.1103/PhysRevLett.110.252001} {\bibfield  {journal}
  {\bibinfo  {journal} {Phys. Rev. Lett.}\ }\textbf {\bibinfo {volume} {110}},\
  \bibinfo {pages} {252001} (\bibinfo {year} {2013})},\ \Eprint
  {https://arxiv.org/abs/1303.5949} {arxiv:1303.5949 [hep-ex]} \BibitemShut
  {NoStop}%
\bibitem [{\citenamefont {Liu}\ \emph {et~al.}(2013)\citenamefont {Liu},
  \citenamefont {Shen}, \citenamefont {Yuan} \emph
  {et~al.}}]{liu2013PRL110-252002}%
  \BibitemOpen
  \bibfield  {author} {\bibinfo {author} {\bibfnamefont {Z.}~\bibnamefont
  {Liu}}, \bibinfo {author} {\bibfnamefont {C.}~\bibnamefont {Shen}}, \bibinfo
  {author} {\bibfnamefont {C.}~\bibnamefont {Yuan}}, \emph {et~al.} (\bibinfo
  {collaboration} {Belle}),\ }\bibfield  {title} {\bibinfo {title} {Study of
  ${e}^{\mathbf{+}}{e}^{\mathbf{\ensuremath{-}}}\ensuremath{\rightarrow}{\ensuremath{\pi}}^{\mathbf{+}}{\ensuremath{\pi}}^{\mathbf{\ensuremath{-}}}{J}/\ensuremath{\psi}$
  and observation of a charged charmoniumlike state at belle},\ }\href
  {https://doi.org/10.1103/PhysRevLett.110.252002} {\bibfield  {journal}
  {\bibinfo  {journal} {Phys. Rev. Lett.}\ }\textbf {\bibinfo {volume} {110}},\
  \bibinfo {pages} {252002} (\bibinfo {year} {2013})},\ \Eprint
  {https://arxiv.org/abs/1304.0121} {arxiv:1304.0121 [hep-ex]} \BibitemShut
  {NoStop}%
\bibitem [{\citenamefont {Abazov}\ \emph {et~al.}(2018)\citenamefont {Abazov},
  \citenamefont {Abbott}, \citenamefont {Acharya} \emph
  {et~al.}}]{abazov2018PRD98-052010}%
  \BibitemOpen
  \bibfield  {author} {\bibinfo {author} {\bibfnamefont {V.~M.}\ \bibnamefont
  {Abazov}}, \bibinfo {author} {\bibfnamefont {B.~K.}\ \bibnamefont {Abbott}},
  \bibinfo {author} {\bibfnamefont {B.~S.}\ \bibnamefont {Acharya}}, \emph
  {et~al.} (\bibinfo {collaboration} {D0}),\ }\bibfield  {title} {\bibinfo
  {title} {Evidence for ${Z}_{c}^{\ifmmode\pm\else\textpm\fi{}}(3900)$ in
  semi-inclusive decays of $b$-flavored hadrons},\ }\href
  {https://doi.org/10.1103/PhysRevD.98.052010} {\bibfield  {journal} {\bibinfo
  {journal} {Phys. Rev. D}\ }\textbf {\bibinfo {volume} {98}},\ \bibinfo
  {pages} {052010} (\bibinfo {year} {2018})},\ \Eprint
  {https://arxiv.org/abs/1807.00183} {arxiv:1807.00183 [hep-ex]} \BibitemShut
  {NoStop}%
\bibitem [{\citenamefont {Ablikim}\ \emph {et~al.}(2021)\citenamefont
  {Ablikim}, \citenamefont {Achasov}, \citenamefont {Adlarson} \emph
  {et~al.}}]{ablikim2021PRL126-102001}%
  \BibitemOpen
  \bibfield  {author} {\bibinfo {author} {\bibfnamefont {M.}~\bibnamefont
  {Ablikim}}, \bibinfo {author} {\bibfnamefont {M.~N.}\ \bibnamefont
  {Achasov}}, \bibinfo {author} {\bibfnamefont {P.~A.}\ \bibnamefont
  {Adlarson}}, \emph {et~al.} (\bibinfo {collaboration} {BESIII}),\ }\bibfield
  {title} {\bibinfo {title} {Observation of a near-threshold structure in the
  ${K}^{+}$ recoil-mass spectra in
  ${e}^{+}{e}^{\ensuremath{-}}\ensuremath{\rightarrow}{K}^{+}({D}_{s}^{\ensuremath{-}}{D}^{*0}+{D}_{s}^{*\ensuremath{-}}{D}^{0})$},\
  }\href {https://doi.org/10.1103/PhysRevLett.126.102001} {\bibfield  {journal}
  {\bibinfo  {journal} {Phys. Rev. Lett.}\ }\textbf {\bibinfo {volume} {126}},\
  \bibinfo {pages} {102001} (\bibinfo {year} {2021})},\ \Eprint
  {https://arxiv.org/abs/2011.07855} {arxiv:2011.07855 [hep-ex]} \BibitemShut
  {NoStop}%
\bibitem [{\citenamefont {Bondar}\ \emph {et~al.}(2012)\citenamefont {Bondar},
  \citenamefont {Garmash}, \citenamefont {Mizuk} \emph
  {et~al.}}]{bondar2012PRL108-122001}%
  \BibitemOpen
  \bibfield  {author} {\bibinfo {author} {\bibfnamefont {A.}~\bibnamefont
  {Bondar}}, \bibinfo {author} {\bibfnamefont {A.}~\bibnamefont {Garmash}},
  \bibinfo {author} {\bibfnamefont {R.}~\bibnamefont {Mizuk}}, \emph {et~al.}
  (\bibinfo {collaboration} {Belle}),\ }\bibfield  {title} {\bibinfo {title}
  {Observation of two charged bottomoniumlike resonances in
  $\ensuremath{\Upsilon}(5{S})$ decays},\ }\href
  {https://doi.org/10.1103/PhysRevLett.108.122001} {\bibfield  {journal}
  {\bibinfo  {journal} {Phys. Rev. Lett.}\ }\textbf {\bibinfo {volume} {108}},\
  \bibinfo {pages} {122001} (\bibinfo {year} {2012})},\ \Eprint
  {https://arxiv.org/abs/1110.2251} {arxiv:1110.2251 [hep-ex]} \BibitemShut
  {NoStop}%
\bibitem [{\citenamefont {Chen}\ \emph {et~al.}(2016)\citenamefont {Chen},
  \citenamefont {Chen}, \citenamefont {Liu},\ and\ \citenamefont
  {Zhu}}]{chen2016PR639-1}%
  \BibitemOpen
  \bibfield  {author} {\bibinfo {author} {\bibfnamefont {H.-X.}\ \bibnamefont
  {Chen}}, \bibinfo {author} {\bibfnamefont {W.}~\bibnamefont {Chen}}, \bibinfo
  {author} {\bibfnamefont {X.}~\bibnamefont {Liu}},\ and\ \bibinfo {author}
  {\bibfnamefont {S.-L.}\ \bibnamefont {Zhu}},\ }\bibfield  {title} {\bibinfo
  {title} {The hidden-charm pentaquark and tetraquark states},\ }\href
  {https://doi.org/10.1016/j.physrep.2016.05.004} {\bibfield  {journal}
  {\bibinfo  {journal} {Phys. Rep.}\ }\textbf {\bibinfo {volume} {639}},\
  \bibinfo {pages} {1} (\bibinfo {year} {2016})},\ \Eprint
  {https://arxiv.org/abs/1601.02092} {arxiv:1601.02092 [hep-ph]} \BibitemShut
  {NoStop}%
\bibitem [{\citenamefont {Lebed}\ \emph {et~al.}(2017)\citenamefont {Lebed},
  \citenamefont {Mitchell},\ and\ \citenamefont
  {Swanson}}]{lebed2017PPNP93-143}%
  \BibitemOpen
  \bibfield  {author} {\bibinfo {author} {\bibfnamefont {R.~F.}\ \bibnamefont
  {Lebed}}, \bibinfo {author} {\bibfnamefont {R.~E.}\ \bibnamefont
  {Mitchell}},\ and\ \bibinfo {author} {\bibfnamefont {E.~S.}\ \bibnamefont
  {Swanson}},\ }\bibfield  {title} {\bibinfo {title} {Heavy-quark {{QCD}}
  exotica},\ }\href {https://doi.org/10.1016/j.ppnp.2016.11.003} {\bibfield
  {journal} {\bibinfo  {journal} {Prog. Part. Nucl. Phys.}\ }\textbf {\bibinfo
  {volume} {93}},\ \bibinfo {pages} {143} (\bibinfo {year} {2017})},\ \Eprint
  {https://arxiv.org/abs/1610.04528} {arxiv:1610.04528 [hep-ph]} \BibitemShut
  {NoStop}%
\bibitem [{\citenamefont {Guo}\ \emph {et~al.}(2018)\citenamefont {Guo},
  \citenamefont {Hanhart}, \citenamefont {Mei{\ss}ner}, \citenamefont {Wang},
  \citenamefont {Zhao},\ and\ \citenamefont {Zou}}]{guo2018RMP90-015004}%
  \BibitemOpen
  \bibfield  {author} {\bibinfo {author} {\bibfnamefont {F.-K.}\ \bibnamefont
  {Guo}}, \bibinfo {author} {\bibfnamefont {C.}~\bibnamefont {Hanhart}},
  \bibinfo {author} {\bibfnamefont {U.-G.}\ \bibnamefont {Mei{\ss}ner}},
  \bibinfo {author} {\bibfnamefont {Q.}~\bibnamefont {Wang}}, \bibinfo {author}
  {\bibfnamefont {Q.}~\bibnamefont {Zhao}},\ and\ \bibinfo {author}
  {\bibfnamefont {B.-S.}\ \bibnamefont {Zou}},\ }\bibfield  {title} {\bibinfo
  {title} {Hadronic molecules},\ }\href
  {https://doi.org/10.1103/RevModPhys.90.015004} {\bibfield  {journal}
  {\bibinfo  {journal} {Rev. Mod. Phys.}\ }\textbf {\bibinfo {volume} {90}},\
  \bibinfo {pages} {015004} (\bibinfo {year} {2018})},\ \Eprint
  {https://arxiv.org/abs/1705.00141} {arxiv:1705.00141 [hep-ph]} \BibitemShut
  {NoStop}%
\bibitem [{\citenamefont {Brambilla}\ \emph {et~al.}(2020)\citenamefont
  {Brambilla}, \citenamefont {Eidelman}, \citenamefont {Hanhart}, \citenamefont
  {Nefediev}, \citenamefont {Shen}, \citenamefont {Thomas}, \citenamefont
  {Vairo},\ and\ \citenamefont {Yuan}}]{brambilla2020PR873-1}%
  \BibitemOpen
  \bibfield  {author} {\bibinfo {author} {\bibfnamefont {N.}~\bibnamefont
  {Brambilla}}, \bibinfo {author} {\bibfnamefont {S.}~\bibnamefont {Eidelman}},
  \bibinfo {author} {\bibfnamefont {C.}~\bibnamefont {Hanhart}}, \bibinfo
  {author} {\bibfnamefont {A.}~\bibnamefont {Nefediev}}, \bibinfo {author}
  {\bibfnamefont {C.-P.}\ \bibnamefont {Shen}}, \bibinfo {author}
  {\bibfnamefont {C.~E.}\ \bibnamefont {Thomas}}, \bibinfo {author}
  {\bibfnamefont {A.}~\bibnamefont {Vairo}},\ and\ \bibinfo {author}
  {\bibfnamefont {C.-Z.}\ \bibnamefont {Yuan}},\ }\bibfield  {title} {\bibinfo
  {title} {The {{XYZ}} states: {{Experimental}} and theoretical status and
  perspectives},\ }\href {https://doi.org/10.1016/j.physrep.2020.05.001}
  {\bibfield  {journal} {\bibinfo  {journal} {Phys. Rep.}\ }\textbf {\bibinfo
  {volume} {873}},\ \bibinfo {pages} {1} (\bibinfo {year} {2020})},\ \Eprint
  {https://arxiv.org/abs/1907.07583} {arxiv:1907.07583 [hep-ex]} \BibitemShut
  {NoStop}%
\bibitem [{\citenamefont {Kalashnikova}\ and\ \citenamefont
  {Nefediev}(2019)}]{kalashnikova2019P62-568}%
  \BibitemOpen
  \bibfield  {author} {\bibinfo {author} {\bibfnamefont {Y.~S.}\ \bibnamefont
  {Kalashnikova}}\ and\ \bibinfo {author} {\bibfnamefont {A.~V.}\ \bibnamefont
  {Nefediev}},\ }\bibfield  {title} {\bibinfo {title} {${X}(3872)$ in the
  molecular model},\ }\href {https://doi.org/10.3367/UFNe.2018.08.038411}
  {\bibfield  {journal} {\bibinfo  {journal} {Phys.-Uspekhi}\ }\textbf
  {\bibinfo {volume} {62}},\ \bibinfo {pages} {568} (\bibinfo {year} {2019})},\
  \Eprint {https://arxiv.org/abs/1811.01324} {arxiv:1811.01324 [hep-ph]}
  \BibitemShut {NoStop}%
\bibitem [{\citenamefont {Meng}\ \emph {et~al.}(2023)\citenamefont {Meng},
  \citenamefont {Wang}, \citenamefont {Wang},\ and\ \citenamefont
  {Zhu}}]{meng2023PR1019-2266}%
  \BibitemOpen
  \bibfield  {author} {\bibinfo {author} {\bibfnamefont {L.}~\bibnamefont
  {Meng}}, \bibinfo {author} {\bibfnamefont {B.}~\bibnamefont {Wang}}, \bibinfo
  {author} {\bibfnamefont {G.-J.}\ \bibnamefont {Wang}},\ and\ \bibinfo
  {author} {\bibfnamefont {S.-L.}\ \bibnamefont {Zhu}},\ }\bibfield  {title}
  {\bibinfo {title} {Chiral perturbation theory for heavy hadrons and chiral
  effective field theory for heavy hadronic molecules},\ }\href
  {https://doi.org/10.1016/j.physrep.2023.04.003} {\bibfield  {journal}
  {\bibinfo  {journal} {Phys. Rep.}\ }\textbf {\bibinfo {volume} {1019}},\
  \bibinfo {pages} {2266} (\bibinfo {year} {2023})},\ \Eprint
  {https://arxiv.org/abs/2204.08716} {arxiv:2204.08716 [hep-ph]} \BibitemShut
  {NoStop}%
\bibitem [{\citenamefont {Abazov}\ \emph {et~al.}(2004)\citenamefont {Abazov},
  \citenamefont {Abbott}, \citenamefont {Abolins} \emph
  {et~al.}}]{abazov2004PRL93-162002}%
  \BibitemOpen
  \bibfield  {author} {\bibinfo {author} {\bibfnamefont {V.}~\bibnamefont
  {Abazov}}, \bibinfo {author} {\bibfnamefont {B.}~\bibnamefont {Abbott}},
  \bibinfo {author} {\bibfnamefont {M.}~\bibnamefont {Abolins}}, \emph {et~al.}
  (\bibinfo {collaboration} {D0}),\ }\bibfield  {title} {\bibinfo {title}
  {Observation and properties of the ${X}(3872)$ decaying to
  ${J}/\ensuremath{\psi}{\ensuremath{\pi}}^{+}{\ensuremath{\pi}}^{\ensuremath{-}}$
  in $p\overline{p}$ collisions at $\sqrt{s}=1.96\text{ }\text{
  }\mathrm{T}\mathrm{e}\mathrm{V}$},\ }\href
  {https://doi.org/10.1103/PhysRevLett.93.162002} {\bibfield  {journal}
  {\bibinfo  {journal} {Phys. Rev. Lett.}\ }\textbf {\bibinfo {volume} {93}},\
  \bibinfo {pages} {162002} (\bibinfo {year} {2004})},\ \Eprint
  {https://arxiv.org/abs/hep-ex/0405004} {arxiv:hep-ex/0405004} \BibitemShut
  {NoStop}%
\bibitem [{\citenamefont {Acosta}\ \emph {et~al.}(2004)\citenamefont {Acosta},
  \citenamefont {Affolder}, \citenamefont {Ahn} \emph
  {et~al.}}]{acosta2004PRL93-072001}%
  \BibitemOpen
  \bibfield  {author} {\bibinfo {author} {\bibfnamefont {D.}~\bibnamefont
  {Acosta}}, \bibinfo {author} {\bibfnamefont {T.}~\bibnamefont {Affolder}},
  \bibinfo {author} {\bibfnamefont {M.}~\bibnamefont {Ahn}}, \emph {et~al.}
  (\bibinfo {collaboration} {CDF}),\ }\bibfield  {title} {\bibinfo {title}
  {Observation of the narrow state
  ${X}(3872)\ensuremath{\rightarrow}{J}/\ensuremath{\psi}{\ensuremath{\pi}}^{+}{\ensuremath{\pi}}^{\ensuremath{-}}$
  in $\overline{p}p$ collisions at $\sqrt{s}=1.96\text{ }\text{
  }\mathrm{T}\mathrm{e}\mathrm{V}$},\ }\href
  {https://doi.org/10.1103/PhysRevLett.93.072001} {\bibfield  {journal}
  {\bibinfo  {journal} {Phys. Rev. Lett.}\ }\textbf {\bibinfo {volume} {93}},\
  \bibinfo {pages} {072001} (\bibinfo {year} {2004})},\ \Eprint
  {https://arxiv.org/abs/hep-ex/0312021} {arxiv:hep-ex/0312021} \BibitemShut
  {NoStop}%
\bibitem [{\citenamefont {Aubert}\ \emph {et~al.}(2005)\citenamefont {Aubert},
  \citenamefont {Barate}, \citenamefont {Boutigny} \emph
  {et~al.}}]{aubert2005PRD71-071103}%
  \BibitemOpen
  \bibfield  {author} {\bibinfo {author} {\bibfnamefont {B.}~\bibnamefont
  {Aubert}}, \bibinfo {author} {\bibfnamefont {R.}~\bibnamefont {Barate}},
  \bibinfo {author} {\bibfnamefont {D.}~\bibnamefont {Boutigny}}, \emph
  {et~al.} (\bibinfo {collaboration} {BaBar}),\ }\bibfield  {title} {\bibinfo
  {title} {Study of the
  ${B}^{\ensuremath{-}}\ensuremath{\rightarrow}{J}/\ensuremath{\psi}{K}^{\ensuremath{-}}{\ensuremath{\pi}}^{+}{\ensuremath{\pi}}^{\ensuremath{-}}$
  decay and measurement of the
  ${B}^{\ensuremath{-}}\ensuremath{\rightarrow}{X}(3872){K}^{\ensuremath{-}}$
  branching fraction},\ }\href {https://doi.org/10.1103/PhysRevD.71.071103}
  {\bibfield  {journal} {\bibinfo  {journal} {Phys. Rev. D}\ }\textbf {\bibinfo
  {volume} {71}},\ \bibinfo {pages} {071103} (\bibinfo {year} {2005})},\
  \Eprint {https://arxiv.org/abs/hep-ex/0406022} {arxiv:hep-ex/0406022}
  \BibitemShut {NoStop}%
\bibitem [{\citenamefont {Workman}\ \emph {et~al.}(2022)\citenamefont
  {Workman}, \citenamefont {Burkert}, \citenamefont {Crede} \emph
  {et~al.}}]{workman2022PTEP2022-083C01}%
  \BibitemOpen
  \bibfield  {author} {\bibinfo {author} {\bibfnamefont {R.}~\bibnamefont
  {Workman}}, \bibinfo {author} {\bibfnamefont {V.}~\bibnamefont {Burkert}},
  \bibinfo {author} {\bibfnamefont {V.}~\bibnamefont {Crede}}, \emph {et~al.}
  (\bibinfo {collaboration} {Particle Data Group}),\ }\bibfield  {title}
  {\bibinfo {title} {Review of {{Particle Physics}}},\ }\href
  {https://doi.org/10.1093/ptep/ptac097} {\bibfield  {journal} {\bibinfo
  {journal} {Prog. Theor. Exp. Phys.}\ }\textbf {\bibinfo {volume} {2022}},\
  \bibinfo {pages} {083C01} (\bibinfo {year} {2022})}\BibitemShut {NoStop}%
\bibitem [{\citenamefont {Aaij}\ \emph {et~al.}(2013)\citenamefont {Aaij},
  \citenamefont {Abellan~Beteta}, \citenamefont {Adeva} \emph
  {et~al.}}]{aaij2013PRL110-222001}%
  \BibitemOpen
  \bibfield  {author} {\bibinfo {author} {\bibfnamefont {R.}~\bibnamefont
  {Aaij}}, \bibinfo {author} {\bibfnamefont {C.}~\bibnamefont
  {Abellan~Beteta}}, \bibinfo {author} {\bibfnamefont {B.}~\bibnamefont
  {Adeva}}, \emph {et~al.} (\bibinfo {collaboration} {LHCb}),\ }\bibfield
  {title} {\bibinfo {title} {Determination of the ${{X}}(3872)$ {{Meson Quantum
  Numbers}}},\ }\href {https://doi.org/10.1103/PhysRevLett.110.222001}
  {\bibfield  {journal} {\bibinfo  {journal} {Phys. Rev. Lett.}\ }\textbf
  {\bibinfo {volume} {110}},\ \bibinfo {pages} {222001} (\bibinfo {year}
  {2013})},\ \Eprint {https://arxiv.org/abs/1302.6269} {arxiv:1302.6269
  [hep-ex]} \BibitemShut {NoStop}%
\bibitem [{\citenamefont {Aaij}\ \emph {et~al.}(2015)\citenamefont {Aaij},
  \citenamefont {Adeva}, \citenamefont {Adinolfi} \emph
  {et~al.}}]{aaij2015PRD92-011102}%
  \BibitemOpen
  \bibfield  {author} {\bibinfo {author} {\bibfnamefont {R.}~\bibnamefont
  {Aaij}}, \bibinfo {author} {\bibfnamefont {B.}~\bibnamefont {Adeva}},
  \bibinfo {author} {\bibfnamefont {M.}~\bibnamefont {Adinolfi}}, \emph
  {et~al.} (\bibinfo {collaboration} {LHCb}),\ }\bibfield  {title} {\bibinfo
  {title} {Quantum numbers of the ${X}(3872)$ state and orbital angular
  momentum in its ${\ensuremath{\rho}}^{0}{J}/\ensuremath{\psi}$ decay},\
  }\href {https://doi.org/10.1103/PhysRevD.92.011102} {\bibfield  {journal}
  {\bibinfo  {journal} {Phys. Rev. D}\ }\textbf {\bibinfo {volume} {92}},\
  \bibinfo {pages} {011102} (\bibinfo {year} {2015})},\ \Eprint
  {https://arxiv.org/abs/1504.06339} {arxiv:1504.06339 [hep-ex]} \BibitemShut
  {NoStop}%
\bibitem [{\citenamefont {Fleming}\ and\ \citenamefont
  {Mehen}(2008)}]{fleming2008PRD78-094019}%
  \BibitemOpen
  \bibfield  {author} {\bibinfo {author} {\bibfnamefont {S.}~\bibnamefont
  {Fleming}}\ and\ \bibinfo {author} {\bibfnamefont {T.}~\bibnamefont
  {Mehen}},\ }\bibfield  {title} {\bibinfo {title} {Hadronic decays of the
  ${X}(3872)$ to ${\ensuremath{\chi}}_{cJ}$ in effective field theory},\ }\href
  {https://doi.org/10.1103/PhysRevD.78.094019} {\bibfield  {journal} {\bibinfo
  {journal} {Phys. Rev. D}\ }\textbf {\bibinfo {volume} {78}},\ \bibinfo
  {pages} {094019} (\bibinfo {year} {2008})},\ \Eprint
  {https://arxiv.org/abs/0807.2674} {arxiv:0807.2674 [hep-ph]} \BibitemShut
  {NoStop}%
\bibitem [{\citenamefont {Dubynskiy}\ and\ \citenamefont
  {Voloshin}(2008)}]{dubynskiy2008PRD77-014013}%
  \BibitemOpen
  \bibfield  {author} {\bibinfo {author} {\bibfnamefont {S.}~\bibnamefont
  {Dubynskiy}}\ and\ \bibinfo {author} {\bibfnamefont {M.~B.}\ \bibnamefont
  {Voloshin}},\ }\bibfield  {title} {\bibinfo {title} {Pionic transitions from
  ${X}(3872)$ to ${\ensuremath{\chi}}_{cJ}$},\ }\href
  {https://doi.org/10.1103/PhysRevD.77.014013} {\bibfield  {journal} {\bibinfo
  {journal} {Phys. Rev. D}\ }\textbf {\bibinfo {volume} {77}},\ \bibinfo
  {pages} {014013} (\bibinfo {year} {2008})},\ \Eprint
  {https://arxiv.org/abs/0709.4474} {arxiv:0709.4474 [hep-ph]} \BibitemShut
  {NoStop}%
\bibitem [{\citenamefont {Guo}\ \emph {et~al.}(2013{\natexlab{a}})\citenamefont
  {Guo}, \citenamefont {Hanhart}, \citenamefont {Mei{\ss}ner}, \citenamefont
  {Wang},\ and\ \citenamefont {Zhao}}]{guo2013PLB725-127}%
  \BibitemOpen
  \bibfield  {author} {\bibinfo {author} {\bibfnamefont {F.-K.}\ \bibnamefont
  {Guo}}, \bibinfo {author} {\bibfnamefont {C.}~\bibnamefont {Hanhart}},
  \bibinfo {author} {\bibfnamefont {U.-G.}\ \bibnamefont {Mei{\ss}ner}},
  \bibinfo {author} {\bibfnamefont {Q.}~\bibnamefont {Wang}},\ and\ \bibinfo
  {author} {\bibfnamefont {Q.}~\bibnamefont {Zhao}},\ }\bibfield  {title}
  {\bibinfo {title} {Production of the ${{X}}(3872)$ in charmonia radiative
  decays},\ }\href {https://doi.org/10.1016/j.physletb.2013.06.053} {\bibfield
  {journal} {\bibinfo  {journal} {Phys. Lett. B}\ }\textbf {\bibinfo {volume}
  {725}},\ \bibinfo {pages} {127} (\bibinfo {year} {2013}{\natexlab{a}})},\
  \Eprint {https://arxiv.org/abs/1306.3096} {arxiv:1306.3096 [hep-ph]}
  \BibitemShut {NoStop}%
\bibitem [{\citenamefont {Wang}\ and\ \citenamefont
  {Huang}(2014{\natexlab{a}})}]{wang2014EPJC74-2891}%
  \BibitemOpen
  \bibfield  {author} {\bibinfo {author} {\bibfnamefont {Z.-G.}\ \bibnamefont
  {Wang}}\ and\ \bibinfo {author} {\bibfnamefont {T.}~\bibnamefont {Huang}},\
  }\bibfield  {title} {\bibinfo {title} {Possible assignments of the
  ${X}(3872)$, ${Z}_c(3900)$, and ${Z}_b(10610)$ as axial-vector molecular
  states},\ }\href {https://doi.org/10.1140/epjc/s10052-014-2891-6} {\bibfield
  {journal} {\bibinfo  {journal} {Eur. Phys. J. C}\ }\textbf {\bibinfo {volume}
  {74}},\ \bibinfo {pages} {2891} (\bibinfo {year} {2014}{\natexlab{a}})},\
  \Eprint {https://arxiv.org/abs/1312.7489} {arxiv:1312.7489 [hep-ph]}
  \BibitemShut {NoStop}%
\bibitem [{\citenamefont {Guo}\ \emph {et~al.}(2015)\citenamefont {Guo},
  \citenamefont {Hanhart}, \citenamefont {Kalashnikova}, \citenamefont
  {Mei{\ss}ner},\ and\ \citenamefont {Nefediev}}]{guo2015PLB742-394}%
  \BibitemOpen
  \bibfield  {author} {\bibinfo {author} {\bibfnamefont {F.-K.}\ \bibnamefont
  {Guo}}, \bibinfo {author} {\bibfnamefont {C.}~\bibnamefont {Hanhart}},
  \bibinfo {author} {\bibfnamefont {{\relax Yu}.~S.}\ \bibnamefont
  {Kalashnikova}}, \bibinfo {author} {\bibfnamefont {U.-G.}\ \bibnamefont
  {Mei{\ss}ner}},\ and\ \bibinfo {author} {\bibfnamefont {A.~V.}\ \bibnamefont
  {Nefediev}},\ }\bibfield  {title} {\bibinfo {title} {What can radiative
  decays of the ${{X}}(3872)$ teach us about its nature?},\ }\href
  {https://doi.org/10.1016/j.physletb.2015.02.013} {\bibfield  {journal}
  {\bibinfo  {journal} {Phys. Lett. B}\ }\textbf {\bibinfo {volume} {742}},\
  \bibinfo {pages} {394} (\bibinfo {year} {2015})},\ \Eprint
  {https://arxiv.org/abs/1410.6712} {arxiv:1410.6712 [hep-ph]} \BibitemShut
  {NoStop}%
\bibitem [{\citenamefont {Mehen}(2015)}]{mehen2015PRD92-034019}%
  \BibitemOpen
  \bibfield  {author} {\bibinfo {author} {\bibfnamefont {T.}~\bibnamefont
  {Mehen}},\ }\bibfield  {title} {\bibinfo {title} {Hadronic loops versus
  factorization in effective field theory calculations of
  ${X}(3872)\ensuremath{\rightarrow}{\ensuremath{\chi}}_{cJ}{\ensuremath{\pi}}^{0}$},\
  }\href {https://doi.org/10.1103/PhysRevD.92.034019} {\bibfield  {journal}
  {\bibinfo  {journal} {Phys. Rev. D}\ }\textbf {\bibinfo {volume} {92}},\
  \bibinfo {pages} {034019} (\bibinfo {year} {2015})},\ \Eprint
  {https://arxiv.org/abs/1503.02719} {arxiv:1503.02719 [hep-ph]} \BibitemShut
  {NoStop}%
\bibitem [{\citenamefont {Dai}\ \emph {et~al.}(2020)\citenamefont {Dai},
  \citenamefont {Guo},\ and\ \citenamefont {Mehen}}]{dai2020PRD101-054024}%
  \BibitemOpen
  \bibfield  {author} {\bibinfo {author} {\bibfnamefont {L.}~\bibnamefont
  {Dai}}, \bibinfo {author} {\bibfnamefont {F.-K.}\ \bibnamefont {Guo}},\ and\
  \bibinfo {author} {\bibfnamefont {T.}~\bibnamefont {Mehen}},\ }\bibfield
  {title} {\bibinfo {title} {Revisiting
  ${X}\mathbf{(}3872\mathbf{)}\ensuremath{\rightarrow}{D}^{0}{{\overline{D}}}^{0}{\ensuremath{\pi}}^{0}$
  in an effective field theory for the ${X}\mathbf{(}3872\mathbf{)}$},\ }\href
  {https://doi.org/10.1103/PhysRevD.101.054024} {\bibfield  {journal} {\bibinfo
   {journal} {Phys. Rev. D}\ }\textbf {\bibinfo {volume} {101}},\ \bibinfo
  {pages} {054024} (\bibinfo {year} {2020})},\ \Eprint
  {https://arxiv.org/abs/1912.04317} {arxiv:1912.04317 [hep-ph]} \BibitemShut
  {NoStop}%
\bibitem [{\citenamefont {Wang}\ \emph
  {et~al.}(2022{\natexlab{a}})\citenamefont {Wang}, \citenamefont {Wu},
  \citenamefont {Li}, \citenamefont {Qin}, \citenamefont {Liu}, \citenamefont
  {An},\ and\ \citenamefont {Xie}}]{wang2022PRD106-074015}%
  \BibitemOpen
  \bibfield  {author} {\bibinfo {author} {\bibfnamefont {Y.}~\bibnamefont
  {Wang}}, \bibinfo {author} {\bibfnamefont {Q.}~\bibnamefont {Wu}}, \bibinfo
  {author} {\bibfnamefont {G.}~\bibnamefont {Li}}, \bibinfo {author}
  {\bibfnamefont {W.-H.}\ \bibnamefont {Qin}}, \bibinfo {author} {\bibfnamefont
  {X.-H.}\ \bibnamefont {Liu}}, \bibinfo {author} {\bibfnamefont {C.-S.}\
  \bibnamefont {An}},\ and\ \bibinfo {author} {\bibfnamefont {J.-J.}\
  \bibnamefont {Xie}},\ }\bibfield  {title} {\bibinfo {title} {Investigations
  of charmless decays of ${{X}}(3872)$ via intermediate meson loops},\ }\href
  {https://doi.org/10.1103/PhysRevD.106.074015} {\bibfield  {journal} {\bibinfo
   {journal} {Phys. Rev. D}\ }\textbf {\bibinfo {volume} {106}},\ \bibinfo
  {pages} {074015} (\bibinfo {year} {2022}{\natexlab{a}})},\ \Eprint
  {https://arxiv.org/abs/2209.12206} {arxiv:2209.12206 [hep-ph]} \BibitemShut
  {NoStop}%
\bibitem [{\citenamefont {Wu}\ \emph {et~al.}(2021)\citenamefont {Wu},
  \citenamefont {Chen},\ and\ \citenamefont {Matsuki}}]{wu2021EPJC81-193}%
  \BibitemOpen
  \bibfield  {author} {\bibinfo {author} {\bibfnamefont {Q.}~\bibnamefont
  {Wu}}, \bibinfo {author} {\bibfnamefont {D.-Y.}\ \bibnamefont {Chen}},\ and\
  \bibinfo {author} {\bibfnamefont {T.}~\bibnamefont {Matsuki}},\ }\bibfield
  {title} {\bibinfo {title} {A phenomenological analysis on isospin-violating
  decay of ${{X}}(3872)$},\ }\href
  {https://doi.org/10.1140/epjc/s10052-021-08984-2} {\bibfield  {journal}
  {\bibinfo  {journal} {Eur. Phys. J. C}\ }\textbf {\bibinfo {volume} {81}},\
  \bibinfo {pages} {193} (\bibinfo {year} {2021})},\ \Eprint
  {https://arxiv.org/abs/2102.08637} {arxiv:2102.08637 [hep-ph]} \BibitemShut
  {NoStop}%
\bibitem [{\citenamefont {Meng}\ \emph {et~al.}(2021)\citenamefont {Meng},
  \citenamefont {Wang}, \citenamefont {Wang},\ and\ \citenamefont
  {Zhu}}]{meng2021PRD104-094003}%
  \BibitemOpen
  \bibfield  {author} {\bibinfo {author} {\bibfnamefont {L.}~\bibnamefont
  {Meng}}, \bibinfo {author} {\bibfnamefont {G.-J.}\ \bibnamefont {Wang}},
  \bibinfo {author} {\bibfnamefont {B.}~\bibnamefont {Wang}},\ and\ \bibinfo
  {author} {\bibfnamefont {S.-L.}\ \bibnamefont {Zhu}},\ }\bibfield  {title}
  {\bibinfo {title} {Revisit the isospin violating decays of ${{X}}(3872)$},\
  }\href {https://doi.org/10.1103/PhysRevD.104.094003} {\bibfield  {journal}
  {\bibinfo  {journal} {Phys. Rev. D}\ }\textbf {\bibinfo {volume} {104}},\
  \bibinfo {pages} {094003} (\bibinfo {year} {2021})},\ \Eprint
  {https://arxiv.org/abs/2109.01333} {arxiv:2109.01333 [hep-ph]} \BibitemShut
  {NoStop}%
\bibitem [{\citenamefont {Chen}(2022)}]{chen2022CTP74-025201}%
  \BibitemOpen
  \bibfield  {author} {\bibinfo {author} {\bibfnamefont {H.-X.}\ \bibnamefont
  {Chen}},\ }\bibfield  {title} {\bibinfo {title} {Decay properties of the
  {{X}}(3872) through the {{Fierz}} rearrangement},\ }\href
  {https://doi.org/10.1088/1572-9494/ac449e} {\bibfield  {journal} {\bibinfo
  {journal} {Commun. Theor. Phys.}\ }\textbf {\bibinfo {volume} {74}},\
  \bibinfo {pages} {025201} (\bibinfo {year} {2022})},\ \Eprint
  {https://arxiv.org/abs/1911.00510} {arxiv:1911.00510 [hep-ph]} \BibitemShut
  {NoStop}%
\bibitem [{\citenamefont {Shi}\ \emph {et~al.}(2021)\citenamefont {Shi},
  \citenamefont {Huang},\ and\ \citenamefont {Wang}}]{shi2021PRD103-094038}%
  \BibitemOpen
  \bibfield  {author} {\bibinfo {author} {\bibfnamefont {P.-P.}\ \bibnamefont
  {Shi}}, \bibinfo {author} {\bibfnamefont {F.}~\bibnamefont {Huang}},\ and\
  \bibinfo {author} {\bibfnamefont {W.-L.}\ \bibnamefont {Wang}},\ }\bibfield
  {title} {\bibinfo {title} {Hidden charm tetraquark states in a diquark
  model},\ }\href {https://doi.org/10.1103/PhysRevD.103.094038} {\bibfield
  {journal} {\bibinfo  {journal} {Phys. Rev. D}\ }\textbf {\bibinfo {volume}
  {103}},\ \bibinfo {pages} {094038} (\bibinfo {year} {2021})},\ \Eprint
  {https://arxiv.org/abs/2105.02397} {arxiv:2105.02397 [hep-ph]} \BibitemShut
  {NoStop}%
\bibitem [{\citenamefont {Maiani}\ \emph {et~al.}(2014)\citenamefont {Maiani},
  \citenamefont {Piccinini}, \citenamefont {Polosa},\ and\ \citenamefont
  {Riquer}}]{maiani2014PRD89-114010}%
  \BibitemOpen
  \bibfield  {author} {\bibinfo {author} {\bibfnamefont {L.}~\bibnamefont
  {Maiani}}, \bibinfo {author} {\bibfnamefont {F.}~\bibnamefont {Piccinini}},
  \bibinfo {author} {\bibfnamefont {A.~D.}\ \bibnamefont {Polosa}},\ and\
  \bibinfo {author} {\bibfnamefont {V.}~\bibnamefont {Riquer}},\ }\bibfield
  {title} {\bibinfo {title} {${{Z}}(4430)$ and a new paradigm for spin
  interactions in tetraquarks},\ }\href
  {https://doi.org/10.1103/PhysRevD.89.114010} {\bibfield  {journal} {\bibinfo
  {journal} {Phys. Rev. D}\ }\textbf {\bibinfo {volume} {89}},\ \bibinfo
  {pages} {114010} (\bibinfo {year} {2014})},\ \Eprint
  {https://arxiv.org/abs/1405.1551} {arxiv:1405.1551 [hep-ph]} \BibitemShut
  {NoStop}%
\bibitem [{\citenamefont {Barnea}\ \emph {et~al.}(2006)\citenamefont {Barnea},
  \citenamefont {Vijande},\ and\ \citenamefont
  {Valcarce}}]{barnea2006PRD73-054004}%
  \BibitemOpen
  \bibfield  {author} {\bibinfo {author} {\bibfnamefont {N.}~\bibnamefont
  {Barnea}}, \bibinfo {author} {\bibfnamefont {J.}~\bibnamefont {Vijande}},\
  and\ \bibinfo {author} {\bibfnamefont {A.}~\bibnamefont {Valcarce}},\
  }\bibfield  {title} {\bibinfo {title} {Four-quark spectroscopy within the
  hyperspherical formalism},\ }\href
  {https://doi.org/10.1103/PhysRevD.73.054004} {\bibfield  {journal} {\bibinfo
  {journal} {Phys. Rev. D}\ }\textbf {\bibinfo {volume} {73}},\ \bibinfo
  {pages} {054004} (\bibinfo {year} {2006})},\ \Eprint
  {https://arxiv.org/abs/hep-ph/0604010} {arxiv:hep-ph/0604010} \BibitemShut
  {NoStop}%
\bibitem [{\citenamefont {Maiani}\ \emph {et~al.}(2005)\citenamefont {Maiani},
  \citenamefont {Piccinini}, \citenamefont {Polosa},\ and\ \citenamefont
  {Riquer}}]{maiani2005PRD71-014028}%
  \BibitemOpen
  \bibfield  {author} {\bibinfo {author} {\bibfnamefont {L.}~\bibnamefont
  {Maiani}}, \bibinfo {author} {\bibfnamefont {F.}~\bibnamefont {Piccinini}},
  \bibinfo {author} {\bibfnamefont {A.~D.}\ \bibnamefont {Polosa}},\ and\
  \bibinfo {author} {\bibfnamefont {V.}~\bibnamefont {Riquer}},\ }\bibfield
  {title} {\bibinfo {title} {Diquark-antidiquark states with hidden or open
  charm and the nature of ${{X}}(3872)$},\ }\href
  {https://doi.org/10.1103/PhysRevD.71.014028} {\bibfield  {journal} {\bibinfo
  {journal} {Phys. Rev. D}\ }\textbf {\bibinfo {volume} {71}},\ \bibinfo
  {pages} {014028} (\bibinfo {year} {2005})},\ \Eprint
  {https://arxiv.org/abs/hep-ph/0412098} {arxiv:hep-ph/0412098} \BibitemShut
  {NoStop}%
\bibitem [{\citenamefont {Wang}(2024)}]{wang2024PRD109-014017}%
  \BibitemOpen
  \bibfield  {author} {\bibinfo {author} {\bibfnamefont {Z.-G.}\ \bibnamefont
  {Wang}},\ }\bibfield  {title} {\bibinfo {title} {Decipher the width of the
  ${{X}}(3872)$ via the {{QCD}} sum rules},\ }\href
  {https://doi.org/10.1103/PhysRevD.109.014017} {\bibfield  {journal} {\bibinfo
   {journal} {Phys. Rev. D}\ }\textbf {\bibinfo {volume} {109}},\ \bibinfo
  {pages} {014017} (\bibinfo {year} {2024})},\ \Eprint
  {https://arxiv.org/abs/2310.02030} {arxiv:2310.02030 [hep-ph]} \BibitemShut
  {NoStop}%
\bibitem [{\citenamefont {Wang}\ and\ \citenamefont
  {Huang}(2014{\natexlab{b}})}]{wang2014PRD89-054019}%
  \BibitemOpen
  \bibfield  {author} {\bibinfo {author} {\bibfnamefont {Z.-G.}\ \bibnamefont
  {Wang}}\ and\ \bibinfo {author} {\bibfnamefont {T.}~\bibnamefont {Huang}},\
  }\bibfield  {title} {\bibinfo {title} {Analysis of the ${X}(3872)$,
  ${Z}_c(3900)$, and ${Z}_c(3885)$ as axial-vector tetraquark states with {QCD}
  sum rules},\ }\href {https://doi.org/10.1103/PhysRevD.89.054019} {\bibfield
  {journal} {\bibinfo  {journal} {Phys. Rev. D}\ }\textbf {\bibinfo {volume}
  {89}},\ \bibinfo {pages} {054019} (\bibinfo {year} {2014}{\natexlab{b}})},\
  \Eprint {https://arxiv.org/abs/1310.2422} {arxiv:1310.2422 [hep-ph]}
  \BibitemShut {NoStop}%
\bibitem [{\citenamefont {Close}\ and\ \citenamefont
  {Godfrey}(2003)}]{close2003PLB574-210}%
  \BibitemOpen
  \bibfield  {author} {\bibinfo {author} {\bibfnamefont {F.~E.}\ \bibnamefont
  {Close}}\ and\ \bibinfo {author} {\bibfnamefont {S.}~\bibnamefont
  {Godfrey}},\ }\bibfield  {title} {\bibinfo {title} {Charmonium hybrid
  production in exclusive {{{\emph{B}}}}-meson decays},\ }\href
  {https://doi.org/10.1016/j.physletb.2003.09.011} {\bibfield  {journal}
  {\bibinfo  {journal} {Phys. Lett. B}\ }\textbf {\bibinfo {volume} {574}},\
  \bibinfo {pages} {210} (\bibinfo {year} {2003})},\ \Eprint
  {https://arxiv.org/abs/hep-ph/0305285} {arxiv:hep-ph/0305285} \BibitemShut
  {NoStop}%
\bibitem [{\citenamefont {Li}(2005)}]{li2005PLB605-306}%
  \BibitemOpen
  \bibfield  {author} {\bibinfo {author} {\bibfnamefont {B.~A.}\ \bibnamefont
  {Li}},\ }\bibfield  {title} {\bibinfo {title} {Is ${{X}}(3872)$ a possible
  candidate as a hybrid meson?},\ }\href
  {https://doi.org/10.1016/j.physletb.2004.11.062} {\bibfield  {journal}
  {\bibinfo  {journal} {Phys. Lett. B}\ }\textbf {\bibinfo {volume} {605}},\
  \bibinfo {pages} {306} (\bibinfo {year} {2005})},\ \Eprint
  {https://arxiv.org/abs/hep-ph/0410264} {arxiv:hep-ph/0410264} \BibitemShut
  {NoStop}%
\bibitem [{\citenamefont {Barnes}\ and\ \citenamefont
  {Godfrey}(2004)}]{barnes2004PRD69-054008}%
  \BibitemOpen
  \bibfield  {author} {\bibinfo {author} {\bibfnamefont {T.}~\bibnamefont
  {Barnes}}\ and\ \bibinfo {author} {\bibfnamefont {S.}~\bibnamefont
  {Godfrey}},\ }\bibfield  {title} {\bibinfo {title} {Charmonium options for
  the ${{X}}(3872)$},\ }\href {https://doi.org/10.1103/PhysRevD.69.054008}
  {\bibfield  {journal} {\bibinfo  {journal} {Phys. Rev. D}\ }\textbf {\bibinfo
  {volume} {69}},\ \bibinfo {pages} {054008} (\bibinfo {year} {2004})},\
  \Eprint {https://arxiv.org/abs/hep-ph/0311162} {arxiv:hep-ph/0311162}
  \BibitemShut {NoStop}%
\bibitem [{\citenamefont {Quigg}(2005)}]{quigg2005NPBPS142-87}%
  \BibitemOpen
  \bibfield  {author} {\bibinfo {author} {\bibfnamefont {C.}~\bibnamefont
  {Quigg}},\ }\bibfield  {title} {\bibinfo {title} {The {{Lost Tribes}} of
  {{Charmonium}}},\ }\href {https://doi.org/10.1016/j.nuclphysbps.2005.01.016}
  {\bibfield  {journal} {\bibinfo  {journal} {Nucl. Phys. B Proc. Suppl.}\
  }\textbf {\bibinfo {volume} {142}},\ \bibinfo {pages} {87} (\bibinfo {year}
  {2005})},\ \Eprint {https://arxiv.org/abs/hep-ph/0407124}
  {arxiv:hep-ph/0407124} \BibitemShut {NoStop}%
\bibitem [{\citenamefont {Eichten}\ \emph {et~al.}(2004)\citenamefont
  {Eichten}, \citenamefont {Lane},\ and\ \citenamefont
  {Quigg}}]{eichten2004PRD69-094019}%
  \BibitemOpen
  \bibfield  {author} {\bibinfo {author} {\bibfnamefont {E.~J.}\ \bibnamefont
  {Eichten}}, \bibinfo {author} {\bibfnamefont {K.}~\bibnamefont {Lane}},\ and\
  \bibinfo {author} {\bibfnamefont {C.}~\bibnamefont {Quigg}},\ }\bibfield
  {title} {\bibinfo {title} {Charmonium levels near threshold and the narrow
  state
  ${X}(3872)\ensuremath{\rightarrow}{\ensuremath{\pi}}^{+}{\ensuremath{\pi}}^{\ensuremath{-}}{J}/\ensuremath{\psi}$},\
  }\href {https://doi.org/10.1103/PhysRevD.69.094019} {\bibfield  {journal}
  {\bibinfo  {journal} {Phys. Rev. D}\ }\textbf {\bibinfo {volume} {69}},\
  \bibinfo {pages} {094019} (\bibinfo {year} {2004})},\ \Eprint
  {https://arxiv.org/abs/hep-ph/0401210} {arxiv:hep-ph/0401210} \BibitemShut
  {NoStop}%
\bibitem [{\citenamefont {Suzuki}(2005)}]{suzuki2005PRD72-114013}%
  \BibitemOpen
  \bibfield  {author} {\bibinfo {author} {\bibfnamefont {M.}~\bibnamefont
  {Suzuki}},\ }\bibfield  {title} {\bibinfo {title} {${{X}}(3872)$ boson:
  {{Molecule}} or charmonium},\ }\href
  {https://doi.org/10.1103/PhysRevD.72.114013} {\bibfield  {journal} {\bibinfo
  {journal} {Phys. Rev. D}\ }\textbf {\bibinfo {volume} {72}},\ \bibinfo
  {pages} {114013} (\bibinfo {year} {2005})},\ \Eprint
  {https://arxiv.org/abs/hep-ph/0508258} {arxiv:hep-ph/0508258} \BibitemShut
  {NoStop}%
\bibitem [{\citenamefont {Achasov}\ and\ \citenamefont
  {Shestakov}(2024{\natexlab{a}})}]{achasov2024PRD109-036028}%
  \BibitemOpen
  \bibfield  {author} {\bibinfo {author} {\bibfnamefont {N.~N.}\ \bibnamefont
  {Achasov}}\ and\ \bibinfo {author} {\bibfnamefont {G.~N.}\ \bibnamefont
  {Shestakov}},\ }\bibfield  {title} {\bibinfo {title} {Toward an estimate of
  the amplitude
  ${X(3872)}\ensuremath{\rightarrow}{\ensuremath{\pi}}^{0}{\ensuremath{\chi}}_{c1}(1{P})$},\
  }\href {https://doi.org/10.1103/PhysRevD.109.036028} {\bibfield  {journal}
  {\bibinfo  {journal} {Phys. Rev. D}\ }\textbf {\bibinfo {volume} {109}},\
  \bibinfo {pages} {036028} (\bibinfo {year} {2024}{\natexlab{a}})}\BibitemShut
  {NoStop}%
\bibitem [{\citenamefont {Achasov}\ and\ \citenamefont
  {Shestakov}(2024{\natexlab{b}})}]{achasov2024x-}%
  \BibitemOpen
  \bibfield  {author} {\bibinfo {author} {\bibfnamefont {N.~N.}\ \bibnamefont
  {Achasov}}\ and\ \bibinfo {author} {\bibfnamefont {G.~N.}\ \bibnamefont
  {Shestakov}},\ }\bibfield  {title} {\bibinfo {title} {Tentative estimates of
  $\mathcal{B}({X}(3872)\to\pi^0\pi^0\chi_{c1})$ and
  $\mathcal{B}({X}(3872)\to\pi^+\pi^-\chi_{c1})$},\ }\href
  {https://doi.org/10.1103/PhysRevD.110.016023} {\bibfield  {journal} {\bibinfo
   {journal} {Phys. Rev. D}\ }\textbf {\bibinfo {volume} {110}},\ \bibinfo
  {pages} {016023} (\bibinfo {year} {2024}{\natexlab{b}})},\ \Eprint
  {https://arxiv.org/abs/2405.09228} {arXiv:2405.09228 [hep-ph]} \BibitemShut
  {NoStop}%
\bibitem [{\citenamefont {Nieves}\ and\ \citenamefont
  {Pav{\'o}n~Valderrama}(2012)}]{nieves2012PRD86-056004}%
  \BibitemOpen
  \bibfield  {author} {\bibinfo {author} {\bibfnamefont {J.}~\bibnamefont
  {Nieves}}\ and\ \bibinfo {author} {\bibfnamefont {M.}~\bibnamefont
  {Pav{\'o}n~Valderrama}},\ }\bibfield  {title} {\bibinfo {title} {Heavy quark
  spin symmetry partners of the ${{X}}(3872)$},\ }\href
  {https://doi.org/10.1103/PhysRevD.86.056004} {\bibfield  {journal} {\bibinfo
  {journal} {Phys. Rev. D}\ }\textbf {\bibinfo {volume} {86}},\ \bibinfo
  {pages} {056004} (\bibinfo {year} {2012})},\ \Eprint
  {https://arxiv.org/abs/1204.2790} {arxiv:1204.2790 [hep-ph]} \BibitemShut
  {NoStop}%
\bibitem [{\citenamefont {Peng}\ \emph {et~al.}(2023)\citenamefont {Peng},
  \citenamefont {Yan},\ and\ \citenamefont
  {Pavon~Valderrama}}]{peng2023PRD108-114001}%
  \BibitemOpen
  \bibfield  {author} {\bibinfo {author} {\bibfnamefont {F.-Z.}\ \bibnamefont
  {Peng}}, \bibinfo {author} {\bibfnamefont {M.-J.}\ \bibnamefont {Yan}},\ and\
  \bibinfo {author} {\bibfnamefont {M.}~\bibnamefont {Pavon~Valderrama}},\
  }\bibfield  {title} {\bibinfo {title} {Heavy- and light-flavor symmetry
  partners of the ${T}_{cc}^{+}(3875)$, the ${X}(3872)$, and the ${X}(3960)$
  from light-meson exchange saturation},\ }\href
  {https://doi.org/10.1103/PhysRevD.108.114001} {\bibfield  {journal} {\bibinfo
   {journal} {Phys. Rev. D}\ }\textbf {\bibinfo {volume} {108}},\ \bibinfo
  {pages} {114001} (\bibinfo {year} {2023})},\ \Eprint
  {https://arxiv.org/abs/2304.13515} {arxiv:2304.13515 [hep-ph]} \BibitemShut
  {NoStop}%
\bibitem [{\citenamefont {Guo}\ \emph {et~al.}(2013{\natexlab{b}})\citenamefont
  {Guo}, \citenamefont {{Hidalgo-Duque}}, \citenamefont {Nieves},\ and\
  \citenamefont {Pav{\'o}n~Valderrama}}]{guo2013PRD88-054007}%
  \BibitemOpen
  \bibfield  {author} {\bibinfo {author} {\bibfnamefont {F.-K.}\ \bibnamefont
  {Guo}}, \bibinfo {author} {\bibfnamefont {C.}~\bibnamefont
  {{Hidalgo-Duque}}}, \bibinfo {author} {\bibfnamefont {J.}~\bibnamefont
  {Nieves}},\ and\ \bibinfo {author} {\bibfnamefont {M.}~\bibnamefont
  {Pav{\'o}n~Valderrama}},\ }\bibfield  {title} {\bibinfo {title} {Consequences
  of heavy-quark symmetries for hadronic molecules},\ }\href
  {https://doi.org/10.1103/PhysRevD.88.054007} {\bibfield  {journal} {\bibinfo
  {journal} {Phys. Rev. D}\ }\textbf {\bibinfo {volume} {88}},\ \bibinfo
  {pages} {054007} (\bibinfo {year} {2013}{\natexlab{b}})},\ \Eprint
  {https://arxiv.org/abs/1303.6608} {arxiv:1303.6608 [hep-ph]} \BibitemShut
  {NoStop}%
\bibitem [{\citenamefont {Tornqvist}(2004)}]{tornqvist2004ax-}%
  \BibitemOpen
  \bibfield  {author} {\bibinfo {author} {\bibfnamefont {N.~A.}\ \bibnamefont
  {Tornqvist}},\ }\bibfield  {title} {\bibinfo {title} {Comment on the narrow
  charmonium state of {{Belle}} at 3871.8 {{MeV}} as a deuson},\ }\href@noop {}
  {\bibfield  {journal} {\bibinfo  {journal} {arXiv:hep-ph/0308277}\ }
  (\bibinfo {year} {2004})},\ \Eprint {https://arxiv.org/abs/hep-ph/0308277}
  {arxiv:hep-ph/0308277} \BibitemShut {NoStop}%
\bibitem [{\citenamefont
  {T{\"o}rnqvist}(1994{\natexlab{a}})}]{tornqvist1994ZPCPF61-525}%
  \BibitemOpen
  \bibfield  {author} {\bibinfo {author} {\bibfnamefont {N.~A.}\ \bibnamefont
  {T{\"o}rnqvist}},\ }\bibfield  {title} {\bibinfo {title} {From the deuteron
  to deusons, an analysis of deuteronlike meson-meson bound states},\ }\href
  {https://doi.org/10.1007/BF01413192} {\bibfield  {journal} {\bibinfo
  {journal} {Z. Phys. C Part. Fields}\ }\textbf {\bibinfo {volume} {61}},\
  \bibinfo {pages} {525} (\bibinfo {year} {1994}{\natexlab{a}})},\ \Eprint
  {https://arxiv.org/abs/hep-ph/9310247} {arxiv:hep-ph/9310247} \BibitemShut
  {NoStop}%
\bibitem [{\citenamefont {Mutuk}\ \emph {et~al.}(2018)\citenamefont {Mutuk},
  \citenamefont {Sara{\c c}}, \citenamefont {G{\"u}m{\"u}s},\ and\
  \citenamefont {Ozpineci}}]{mutuk2018EPJC78-904}%
  \BibitemOpen
  \bibfield  {author} {\bibinfo {author} {\bibfnamefont {H.}~\bibnamefont
  {Mutuk}}, \bibinfo {author} {\bibfnamefont {Y.}~\bibnamefont {Sara{\c c}}},
  \bibinfo {author} {\bibfnamefont {H.}~\bibnamefont {G{\"u}m{\"u}s}},\ and\
  \bibinfo {author} {\bibfnamefont {A.}~\bibnamefont {Ozpineci}},\ }\bibfield
  {title} {\bibinfo {title} {${X}(3872)$ and its heavy quark spin symmetry
  partners in {QCD} sum rules},\ }\href
  {https://doi.org/10.1140/epjc/s10052-018-6382-z} {\bibfield  {journal}
  {\bibinfo  {journal} {Eur. Phys. J. C}\ }\textbf {\bibinfo {volume} {78}},\
  \bibinfo {pages} {904} (\bibinfo {year} {2018})},\ \Eprint
  {https://arxiv.org/abs/1807.04091} {arxiv:1807.04091 [hep-ph]} \BibitemShut
  {NoStop}%
\bibitem [{\citenamefont {Wang}(2021)}]{wang2021IJMPA36-2150107}%
  \BibitemOpen
  \bibfield  {author} {\bibinfo {author} {\bibfnamefont {Z.-G.}\ \bibnamefont
  {Wang}},\ }\bibfield  {title} {\bibinfo {title} {Analysis of the
  {{Hidden-charm Tetraquark}} molecule mass spectrum with the {{QCD}} sum
  rules},\ }\href {https://doi.org/10.1142/S0217751X21501074} {\bibfield
  {journal} {\bibinfo  {journal} {Int. J. Mod. Phys. A}\ }\textbf {\bibinfo
  {volume} {36}},\ \bibinfo {pages} {2150107} (\bibinfo {year} {2021})},\
  \Eprint {https://arxiv.org/abs/2012.11869} {arxiv:2012.11869 [hep-ph]}
  \BibitemShut {NoStop}%
\bibitem [{\citenamefont {Wang}(2020)}]{wang2020PRD102-014018}%
  \BibitemOpen
  \bibfield  {author} {\bibinfo {author} {\bibfnamefont {Z.-G.}\ \bibnamefont
  {Wang}},\ }\bibfield  {title} {\bibinfo {title} {Analysis of the hidden-charm
  tetraquark mass spectrum with the {{QCD}} sum rules},\ }\href
  {https://doi.org/10.1103/PhysRevD.102.014018} {\bibfield  {journal} {\bibinfo
   {journal} {Phys. Rev. D}\ }\textbf {\bibinfo {volume} {102}},\ \bibinfo
  {pages} {014018} (\bibinfo {year} {2020})},\ \Eprint
  {https://arxiv.org/abs/1908.07914} {arxiv:1908.07914 [hep-ph]} \BibitemShut
  {NoStop}%
\bibitem [{\citenamefont {Wang}\ \emph
  {et~al.}(2022{\natexlab{b}})\citenamefont {Wang}, \citenamefont {Gao},
  \citenamefont {Zhu} \emph {et~al.}}]{wang2022PRD105-112011}%
  \BibitemOpen
  \bibfield  {author} {\bibinfo {author} {\bibfnamefont {X.}~\bibnamefont
  {Wang}}, \bibinfo {author} {\bibfnamefont {B.}~\bibnamefont {Gao}}, \bibinfo
  {author} {\bibfnamefont {W.}~\bibnamefont {Zhu}}, \emph {et~al.} (\bibinfo
  {collaboration} {Belle}),\ }\bibfield  {title} {\bibinfo {title} {Study of
  $\ensuremath{\gamma}\ensuremath{\gamma}\ensuremath{\rightarrow}\ensuremath{\gamma}\ensuremath{\psi}(2{S})$
  at belle},\ }\href {https://doi.org/10.1103/PhysRevD.105.112011} {\bibfield
  {journal} {\bibinfo  {journal} {Phys. Rev. D}\ }\textbf {\bibinfo {volume}
  {105}},\ \bibinfo {pages} {112011} (\bibinfo {year} {2022}{\natexlab{b}})},\
  \Eprint {https://arxiv.org/abs/2105.06605} {arxiv:2105.06605 [hep-ex]}
  \BibitemShut {NoStop}%
\bibitem [{\citenamefont {Albaladejo}\ \emph {et~al.}(2015)\citenamefont
  {Albaladejo}, \citenamefont {Guo}, \citenamefont {{Hidalgo-Duque}},
  \citenamefont {Nieves},\ and\ \citenamefont
  {Valderrama}}]{albaladejo2015EPJC75-547}%
  \BibitemOpen
  \bibfield  {author} {\bibinfo {author} {\bibfnamefont {M.}~\bibnamefont
  {Albaladejo}}, \bibinfo {author} {\bibfnamefont {F.-K.}\ \bibnamefont {Guo}},
  \bibinfo {author} {\bibfnamefont {C.}~\bibnamefont {{Hidalgo-Duque}}},
  \bibinfo {author} {\bibfnamefont {J.}~\bibnamefont {Nieves}},\ and\ \bibinfo
  {author} {\bibfnamefont {M.~P.}\ \bibnamefont {Valderrama}},\ }\bibfield
  {title} {\bibinfo {title} {Decay widths of the spin-2 partners of the
  ${{X}}(3872)$},\ }\href {https://doi.org/10.1140/epjc/s10052-015-3753-6}
  {\bibfield  {journal} {\bibinfo  {journal} {Eur. Phys. J. C}\ }\textbf
  {\bibinfo {volume} {75}},\ \bibinfo {pages} {547} (\bibinfo {year} {2015})},\
  \Eprint {https://arxiv.org/abs/1504.00861} {arxiv:1504.00861 [hep-ph]}
  \BibitemShut {NoStop}%
\bibitem [{\citenamefont {Shi}\ \emph {et~al.}(2023)\citenamefont {Shi},
  \citenamefont {Dias},\ and\ \citenamefont {Guo}}]{shi2023PLB843-137987}%
  \BibitemOpen
  \bibfield  {author} {\bibinfo {author} {\bibfnamefont {P.-P.}\ \bibnamefont
  {Shi}}, \bibinfo {author} {\bibfnamefont {J.~M.}\ \bibnamefont {Dias}},\ and\
  \bibinfo {author} {\bibfnamefont {F.-K.}\ \bibnamefont {Guo}},\ }\bibfield
  {title} {\bibinfo {title} {Radiative decays of the spin-2 partner of
  ${{X}}(3872)$},\ }\href {https://doi.org/10.1016/j.physletb.2023.137987}
  {\bibfield  {journal} {\bibinfo  {journal} {Phys. Lett. B}\ }\textbf
  {\bibinfo {volume} {843}},\ \bibinfo {pages} {137987} (\bibinfo {year}
  {2023})},\ \Eprint {https://arxiv.org/abs/2302.13017} {arxiv:2302.13017
  [hep-ph]} \BibitemShut {NoStop}%
\bibitem [{\citenamefont {Zheng}\ \emph {et~al.}(2024)\citenamefont {Zheng},
  \citenamefont {Cai}, \citenamefont {Li}, \citenamefont {Liu}, \citenamefont
  {Wu},\ and\ \citenamefont {Wu}}]{zheng2024PRD109-014027}%
  \BibitemOpen
  \bibfield  {author} {\bibinfo {author} {\bibfnamefont {Y.}~\bibnamefont
  {Zheng}}, \bibinfo {author} {\bibfnamefont {Z.}~\bibnamefont {Cai}}, \bibinfo
  {author} {\bibfnamefont {G.}~\bibnamefont {Li}}, \bibinfo {author}
  {\bibfnamefont {S.}~\bibnamefont {Liu}}, \bibinfo {author} {\bibfnamefont
  {J.}~\bibnamefont {Wu}},\ and\ \bibinfo {author} {\bibfnamefont
  {Q.}~\bibnamefont {Wu}},\ }\bibfield  {title} {\bibinfo {title} {Hidden
  charmonium decays of spin-2 partner of ${{X}}(3872)$},\ }\href
  {https://doi.org/10.1103/PhysRevD.109.014027} {\bibfield  {journal} {\bibinfo
   {journal} {Phys. Rev. D}\ }\textbf {\bibinfo {volume} {109}},\ \bibinfo
  {pages} {014027} (\bibinfo {year} {2024})},\ \Eprint
  {https://arxiv.org/abs/2401.03219} {arxiv:2401.03219 [hep-ph]} \BibitemShut
  {NoStop}%
\bibitem [{\citenamefont {Shi}\ \emph {et~al.}(2024)\citenamefont {Shi},
  \citenamefont {Baru}, \citenamefont {Guo}, \citenamefont {Hanhart},\ and\
  \citenamefont {Nefediev}}]{shi2024CPL41-031301}%
  \BibitemOpen
  \bibfield  {author} {\bibinfo {author} {\bibfnamefont {P.-P.}\ \bibnamefont
  {Shi}}, \bibinfo {author} {\bibfnamefont {V.}~\bibnamefont {Baru}}, \bibinfo
  {author} {\bibfnamefont {F.-K.}\ \bibnamefont {Guo}}, \bibinfo {author}
  {\bibfnamefont {C.}~\bibnamefont {Hanhart}},\ and\ \bibinfo {author}
  {\bibfnamefont {A.}~\bibnamefont {Nefediev}},\ }\bibfield  {title} {\bibinfo
  {title} {Production of the ${X}(4014)$ as the spin-2 partner of ${X}(3872)$
  in $e^+e^-$ collisions},\ }\href
  {https://doi.org/10.1088/0256-307X/41/3/031301} {\bibfield  {journal}
  {\bibinfo  {journal} {Chin. Phys. Lett.}\ }\textbf {\bibinfo {volume} {41}},\
  \bibinfo {pages} {031301} (\bibinfo {year} {2024})},\ \Eprint
  {https://arxiv.org/abs/2312.05389} {arxiv:2312.05389 [hep-ph]} \BibitemShut
  {NoStop}%
\bibitem [{\citenamefont {Wu}\ \emph {et~al.}(2024)\citenamefont {Wu},
  \citenamefont {Liu},\ and\ \citenamefont {Geng}}]{wu2024EPJC84-147}%
  \BibitemOpen
  \bibfield  {author} {\bibinfo {author} {\bibfnamefont {Q.}~\bibnamefont
  {Wu}}, \bibinfo {author} {\bibfnamefont {M.-Z.}\ \bibnamefont {Liu}},\ and\
  \bibinfo {author} {\bibfnamefont {L.-S.}\ \bibnamefont {Geng}},\ }\bibfield
  {title} {\bibinfo {title} {Productions of ${{X}}(3872)$, ${{Z}}_c(3900)$,
  ${{X}}_2(4013)$, and ${{Z}}_c(4020)$ in ${B}_{(s)}$ decays offer strong clues
  on their molecular nature},\ }\href
  {https://doi.org/10.1140/epjc/s10052-024-12501-6} {\bibfield  {journal}
  {\bibinfo  {journal} {Eur. Phys. J. C}\ }\textbf {\bibinfo {volume} {84}},\
  \bibinfo {pages} {147} (\bibinfo {year} {2024})},\ \Eprint
  {https://arxiv.org/abs/2304.05269} {arxiv:2304.05269 [hep-ph]} \BibitemShut
  {NoStop}%
\bibitem [{\citenamefont {Liu}\ \emph {et~al.}(2024)\citenamefont {Liu},
  \citenamefont {Ling},\ and\ \citenamefont {Geng}}]{liu2024a[x-}%
  \BibitemOpen
  \bibfield  {author} {\bibinfo {author} {\bibfnamefont {M.-Z.}\ \bibnamefont
  {Liu}}, \bibinfo {author} {\bibfnamefont {X.-Z.}\ \bibnamefont {Ling}},\ and\
  \bibinfo {author} {\bibfnamefont {L.-S.}\ \bibnamefont {Geng}},\ }\bibfield
  {title} {\bibinfo {title} {Productions of ${{X}}(3872)$/${{Z}}_c(3900)$ and
  ${{X}}_2(4013)$/${{Z}}_c(4020)$ in ${{Y}}(4220)$ and ${{Y}}(4360)$ decays},\
  }\bibfield  {journal} {\bibinfo  {journal} {arXiv:2404.07681 [hep-ph]}\
  }\href {https://doi.org/10.48550/arXiv.2404.07681}
  {10.48550/arXiv.2404.07681} (\bibinfo {year} {2024}),\ \Eprint
  {https://arxiv.org/abs/2404.07681} {arxiv:2404.07681 [hep-ph]} \BibitemShut
  {NoStop}%
\bibitem [{\citenamefont {Ablikim}\ \emph {et~al.}(2019)\citenamefont
  {Ablikim}, \citenamefont {Achasov}, \citenamefont {Ahmed} \emph
  {et~al.}}]{ablikim2019PRL122-202001}%
  \BibitemOpen
  \bibfield  {author} {\bibinfo {author} {\bibfnamefont {M.}~\bibnamefont
  {Ablikim}}, \bibinfo {author} {\bibfnamefont {M.}~\bibnamefont {Achasov}},
  \bibinfo {author} {\bibfnamefont {S.}~\bibnamefont {Ahmed}}, \emph {et~al.}
  (\bibinfo {collaboration} {BESIII}),\ }\bibfield  {title} {\bibinfo {title}
  {Observation of the decay ${X}(3872) \to \pi^0 \chi_{c1}(1{P})$},\ }\href
  {https://doi.org/10.1103/PhysRevLett.122.202001} {\bibfield  {journal}
  {\bibinfo  {journal} {Phys. Rev. Lett.}\ }\textbf {\bibinfo {volume} {122}},\
  \bibinfo {pages} {202001} (\bibinfo {year} {2019})},\ \Eprint
  {https://arxiv.org/abs/1901.03992} {arxiv:1901.03992 [hep-ex]} \BibitemShut
  {NoStop}%
\bibitem [{\citenamefont {Ablikim}\ \emph {et~al.}(2024)\citenamefont
  {Ablikim}, \citenamefont {Achasov}, \citenamefont {Adlarson} \emph
  {et~al.}}]{ablikim2024PRD109-L071101}%
  \BibitemOpen
  \bibfield  {author} {\bibinfo {author} {\bibfnamefont {M.}~\bibnamefont
  {Ablikim}}, \bibinfo {author} {\bibfnamefont {M.~N.}\ \bibnamefont
  {Achasov}}, \bibinfo {author} {\bibfnamefont {P.~A.}\ \bibnamefont
  {Adlarson}}, \emph {et~al.} (\bibinfo {collaboration} {BESIII}),\ }\bibfield
  {title} {\bibinfo {title} {Search for the decay
  $\chi_{c1}(3872)\to\pi^{+}\pi^{-}\chi_{c1}$},\ }\href
  {https://doi.org/10.1103/PhysRevD.109.L071101} {\bibfield  {journal}
  {\bibinfo  {journal} {Phys. Rev. D}\ }\textbf {\bibinfo {volume} {109}},\
  \bibinfo {pages} {L071101} (\bibinfo {year} {2024})},\ \Eprint
  {https://arxiv.org/abs/2312.13593} {arxiv:2312.13593 [hep-ex]} \BibitemShut
  {NoStop}%
\bibitem [{\citenamefont
  {T{\"o}rnqvist}(1994{\natexlab{b}})}]{tornqvist1994ZPC61-525}%
  \BibitemOpen
  \bibfield  {author} {\bibinfo {author} {\bibfnamefont {N.~A.}\ \bibnamefont
  {T{\"o}rnqvist}},\ }\bibfield  {title} {\bibinfo {title} {From the deuteron
  to deusons, an analysis of deuteronlike meson-meson bound states},\ }\href
  {https://doi.org/10.1007/BF01413192} {\bibfield  {journal} {\bibinfo
  {journal} {Z. Phys. C}\ }\textbf {\bibinfo {volume} {61}},\ \bibinfo {pages}
  {525} (\bibinfo {year} {1994}{\natexlab{b}})},\ \Eprint
  {https://arxiv.org/abs/hep-ph/9310247} {arxiv:hep-ph/9310247} \BibitemShut
  {NoStop}%
\bibitem [{\citenamefont {{Hidalgo-Duque}}\ \emph {et~al.}(2013)\citenamefont
  {{Hidalgo-Duque}}, \citenamefont {Nieves},\ and\ \citenamefont
  {Valderrama}}]{hidalgo-duque2013PRD87-076006}%
  \BibitemOpen
  \bibfield  {author} {\bibinfo {author} {\bibfnamefont {C.}~\bibnamefont
  {{Hidalgo-Duque}}}, \bibinfo {author} {\bibfnamefont {J.}~\bibnamefont
  {Nieves}},\ and\ \bibinfo {author} {\bibfnamefont {M.~P.}\ \bibnamefont
  {Valderrama}},\ }\bibfield  {title} {\bibinfo {title} {Light flavor and heavy
  quark spin symmetry in heavy meson molecules},\ }\href
  {https://doi.org/10.1103/PhysRevD.87.076006} {\bibfield  {journal} {\bibinfo
  {journal} {Phys. Rev. D}\ }\textbf {\bibinfo {volume} {87}},\ \bibinfo
  {pages} {076006} (\bibinfo {year} {2013})},\ \Eprint
  {https://arxiv.org/abs/1210.5431} {arxiv:1210.5431 [hep-ph]} \BibitemShut
  {NoStop}%
\bibitem [{\citenamefont {Baru}\ \emph {et~al.}(2016)\citenamefont {Baru},
  \citenamefont {Epelbaum}, \citenamefont {Filin}, \citenamefont {Hanhart},
  \citenamefont {Mei{\ss}ner},\ and\ \citenamefont
  {Nefediev}}]{baru2016PLB763-20}%
  \BibitemOpen
  \bibfield  {author} {\bibinfo {author} {\bibfnamefont {V.}~\bibnamefont
  {Baru}}, \bibinfo {author} {\bibfnamefont {E.}~\bibnamefont {Epelbaum}},
  \bibinfo {author} {\bibfnamefont {A.~A.}\ \bibnamefont {Filin}}, \bibinfo
  {author} {\bibfnamefont {C.}~\bibnamefont {Hanhart}}, \bibinfo {author}
  {\bibfnamefont {U.-G.}\ \bibnamefont {Mei{\ss}ner}},\ and\ \bibinfo {author}
  {\bibfnamefont {A.~V.}\ \bibnamefont {Nefediev}},\ }\bibfield  {title}
  {\bibinfo {title} {Heavy-quark spin symmetry partners of the ${{X}}(3872)$
  revisited},\ }\href {https://doi.org/10.1016/j.physletb.2016.10.008}
  {\bibfield  {journal} {\bibinfo  {journal} {Phys. Lett. B}\ }\textbf
  {\bibinfo {volume} {763}},\ \bibinfo {pages} {20} (\bibinfo {year} {2016})},\
  \Eprint {https://arxiv.org/abs/1605.09649} {arxiv:1605.09649 [hep-ph]}
  \BibitemShut {NoStop}%
\bibitem [{\citenamefont {Jia}\ \emph {et~al.}(2024)\citenamefont {Jia},
  \citenamefont {Zhang}, \citenamefont {Qin},\ and\ \citenamefont
  {Li}}]{jia2024PRD109-034017}%
  \BibitemOpen
  \bibfield  {author} {\bibinfo {author} {\bibfnamefont {Z.-S.}\ \bibnamefont
  {Jia}}, \bibinfo {author} {\bibfnamefont {Z.-H.}\ \bibnamefont {Zhang}},
  \bibinfo {author} {\bibfnamefont {W.-H.}\ \bibnamefont {Qin}},\ and\ \bibinfo
  {author} {\bibfnamefont {G.}~\bibnamefont {Li}},\ }\bibfield  {title}
  {\bibinfo {title} {Hunting for ${X}_{b}$ via hidden bottomonium decays
  ${X}_{b}\ensuremath{\rightarrow}\ensuremath{\pi}\ensuremath{\pi}{\ensuremath{\chi}}_{bJ}$},\
  }\href {https://doi.org/10.1103/PhysRevD.109.034017} {\bibfield  {journal}
  {\bibinfo  {journal} {Phys. Rev. D}\ }\textbf {\bibinfo {volume} {109}},\
  \bibinfo {pages} {034017} (\bibinfo {year} {2024})},\ \Eprint
  {https://arxiv.org/abs/2311.15527} {arxiv:2311.15527 [hep-ph]} \BibitemShut
  {NoStop}%
\bibitem [{\citenamefont {Hu}\ and\ \citenamefont
  {Mehen}(2006)}]{hu2006PRD73-054003}%
  \BibitemOpen
  \bibfield  {author} {\bibinfo {author} {\bibfnamefont {J.}~\bibnamefont
  {Hu}}\ and\ \bibinfo {author} {\bibfnamefont {T.}~\bibnamefont {Mehen}},\
  }\bibfield  {title} {\bibinfo {title} {Chiral {{Lagrangian}} with heavy
  quark-diquark symmetry},\ }\href {https://doi.org/10.1103/PhysRevD.73.054003}
  {\bibfield  {journal} {\bibinfo  {journal} {Phys. Rev. D}\ }\textbf {\bibinfo
  {volume} {73}},\ \bibinfo {pages} {054003} (\bibinfo {year} {2006})},\
  \Eprint {https://arxiv.org/abs/hep-ph/0511321} {arxiv:hep-ph/0511321}
  \BibitemShut {NoStop}%
\bibitem [{\citenamefont {Casalbuoni}\ \emph {et~al.}(1997)\citenamefont
  {Casalbuoni}, \citenamefont {Deandrea}, \citenamefont {Di~Bartolomeo},
  \citenamefont {Gatto}, \citenamefont {Feruglio},\ and\ \citenamefont
  {Nardulli}}]{casalbuoni1997PR281-145}%
  \BibitemOpen
  \bibfield  {author} {\bibinfo {author} {\bibfnamefont {R.}~\bibnamefont
  {Casalbuoni}}, \bibinfo {author} {\bibfnamefont {A.}~\bibnamefont
  {Deandrea}}, \bibinfo {author} {\bibfnamefont {N.}~\bibnamefont
  {Di~Bartolomeo}}, \bibinfo {author} {\bibfnamefont {R.}~\bibnamefont
  {Gatto}}, \bibinfo {author} {\bibfnamefont {F.}~\bibnamefont {Feruglio}},\
  and\ \bibinfo {author} {\bibfnamefont {G.}~\bibnamefont {Nardulli}},\
  }\bibfield  {title} {\bibinfo {title} {Phenomenology of heavy meson chiral
  lagrangians},\ }\href {https://doi.org/10.1016/S0370-1573(96)00027-0}
  {\bibfield  {journal} {\bibinfo  {journal} {Phys. Rep.}\ }\textbf {\bibinfo
  {volume} {281}},\ \bibinfo {pages} {145} (\bibinfo {year} {1997})},\ \Eprint
  {https://arxiv.org/abs/hep-ph/9605342} {arxiv:hep-ph/9605342} \BibitemShut
  {NoStop}%
\bibitem [{\citenamefont {Guo}\ \emph {et~al.}(2011)\citenamefont {Guo},
  \citenamefont {Hanhart}, \citenamefont {Li}, \citenamefont {Mei{\ss}ner},\
  and\ \citenamefont {Zhao}}]{guo2011PRD83-034013}%
  \BibitemOpen
  \bibfield  {author} {\bibinfo {author} {\bibfnamefont {F.-K.}\ \bibnamefont
  {Guo}}, \bibinfo {author} {\bibfnamefont {C.}~\bibnamefont {Hanhart}},
  \bibinfo {author} {\bibfnamefont {G.}~\bibnamefont {Li}}, \bibinfo {author}
  {\bibfnamefont {U.-G.}\ \bibnamefont {Mei{\ss}ner}},\ and\ \bibinfo {author}
  {\bibfnamefont {Q.}~\bibnamefont {Zhao}},\ }\bibfield  {title} {\bibinfo
  {title} {Effect of charmed meson loops on charmonium transitions},\ }\href
  {https://doi.org/10.1103/PhysRevD.83.034013} {\bibfield  {journal} {\bibinfo
  {journal} {Phys. Rev. D}\ }\textbf {\bibinfo {volume} {83}},\ \bibinfo
  {pages} {034013} (\bibinfo {year} {2011})},\ \Eprint
  {https://arxiv.org/abs/1008.3632} {arxiv:1008.3632 [hep-ph]} \BibitemShut
  {NoStop}%
\bibitem [{\citenamefont {Zhou}\ \emph {et~al.}(2019)\citenamefont {Zhou},
  \citenamefont {Yu},\ and\ \citenamefont {Xiao}}]{zhou2019PRD100-094025}%
  \BibitemOpen
  \bibfield  {author} {\bibinfo {author} {\bibfnamefont {Z.-Y.}\ \bibnamefont
  {Zhou}}, \bibinfo {author} {\bibfnamefont {M.-T.}\ \bibnamefont {Yu}},\ and\
  \bibinfo {author} {\bibfnamefont {Z.}~\bibnamefont {Xiao}},\ }\bibfield
  {title} {\bibinfo {title} {Decays of ${X}(3872)$ to
  ${\ensuremath{\chi}}_{cJ}{\ensuremath{\pi}}^{0}$ and
  ${J}/\ensuremath{\psi}{\ensuremath{\pi}}^{+}{\ensuremath{\pi}}^{\ensuremath{-}}$},\
  }\href {https://doi.org/10.1103/PhysRevD.100.094025} {\bibfield  {journal}
  {\bibinfo  {journal} {Phys. Rev. D}\ }\textbf {\bibinfo {volume} {100}},\
  \bibinfo {pages} {094025} (\bibinfo {year} {2019})},\ \Eprint
  {https://arxiv.org/abs/1904.07509} {arxiv:1904.07509 [hep-ph]} \BibitemShut
  {NoStop}%
\bibitem [{\citenamefont {Liu}\ \emph {et~al.}(2019)\citenamefont {Liu},
  \citenamefont {Li}, \citenamefont {Xie},\ and\ \citenamefont
  {Zhao}}]{liu2019PRD100-054006}%
  \BibitemOpen
  \bibfield  {author} {\bibinfo {author} {\bibfnamefont {X.-H.}\ \bibnamefont
  {Liu}}, \bibinfo {author} {\bibfnamefont {G.}~\bibnamefont {Li}}, \bibinfo
  {author} {\bibfnamefont {J.-J.}\ \bibnamefont {Xie}},\ and\ \bibinfo {author}
  {\bibfnamefont {Q.}~\bibnamefont {Zhao}},\ }\bibfield  {title} {\bibinfo
  {title} {Visible narrow cusp structure in
  ${\mathrm{\ensuremath{\Lambda}}}_{c}^{+}\ensuremath{\rightarrow}p{K}^{\ensuremath{-}}{\ensuremath{\pi}}^{+}$
  enhanced by triangle singularity},\ }\href
  {https://doi.org/10.1103/PhysRevD.100.054006} {\bibfield  {journal} {\bibinfo
   {journal} {Phys. Rev. D}\ }\textbf {\bibinfo {volume} {100}},\ \bibinfo
  {pages} {054006} (\bibinfo {year} {2019})},\ \Eprint
  {https://arxiv.org/abs/1906.07942} {arxiv:1906.07942 [hep-ph]} \BibitemShut
  {NoStop}%
\bibitem [{\citenamefont {Chen}\ and\ \citenamefont
  {Liu}(2011)}]{chen2011PRD84-094003}%
  \BibitemOpen
  \bibfield  {author} {\bibinfo {author} {\bibfnamefont {D.-Y.}\ \bibnamefont
  {Chen}}\ and\ \bibinfo {author} {\bibfnamefont {X.}~\bibnamefont {Liu}},\
  }\bibfield  {title} {\bibinfo {title} {${Z}_b(10610)$ and ${Z}_b(10650)$
  structures produced by the initial single pion emission in the
  ${\Upsilon}(5{S})$ decays},\ }\href
  {https://doi.org/10.1103/PhysRevD.84.094003} {\bibfield  {journal} {\bibinfo
  {journal} {Phys. Rev. D}\ }\textbf {\bibinfo {volume} {84}},\ \bibinfo
  {pages} {094003} (\bibinfo {year} {2011})},\ \Eprint
  {https://arxiv.org/abs/1106.3798} {arxiv:1106.3798 [hep-ph]} \BibitemShut
  {NoStop}%
\bibitem [{\citenamefont {Bugg}(2011)}]{bugg2011E96-11002}%
  \BibitemOpen
  \bibfield  {author} {\bibinfo {author} {\bibfnamefont {D.~V.}\ \bibnamefont
  {Bugg}},\ }\bibfield  {title} {\bibinfo {title} {An explanation of {{Belle}}
  states ${{Zb}}(10610)$ and ${{Zb}}(10650)$},\ }\href
  {https://doi.org/10.1209/0295-5075/96/11002} {\bibfield  {journal} {\bibinfo
  {journal} {EPL}\ }\textbf {\bibinfo {volume} {96}},\ \bibinfo {pages} {11002}
  (\bibinfo {year} {2011})},\ \Eprint {https://arxiv.org/abs/1105.5492}
  {arxiv:1105.5492 [hep-ph]} \BibitemShut {NoStop}%
\bibitem [{\citenamefont {Xie}\ and\ \citenamefont
  {Oset}(2019)}]{xie2019PLB792-450}%
  \BibitemOpen
  \bibfield  {author} {\bibinfo {author} {\bibfnamefont {J.-J.}\ \bibnamefont
  {Xie}}\ and\ \bibinfo {author} {\bibfnamefont {E.}~\bibnamefont {Oset}},\
  }\bibfield  {title} {\bibinfo {title} {Search for the ${\Sigma}^*$ state in
  ${\Lambda}^+_c \to \pi^+ \pi^0 \pi^-{\Sigma}^+$ decay by triangle
  singularity},\ }\href {https://doi.org/10.1016/j.physletb.2019.04.011}
  {\bibfield  {journal} {\bibinfo  {journal} {Phys. Lett. B}\ }\textbf
  {\bibinfo {volume} {792}},\ \bibinfo {pages} {450} (\bibinfo {year}
  {2019})},\ \Eprint {https://arxiv.org/abs/1811.07247} {arxiv:1811.07247
  [hep-ph]} \BibitemShut {NoStop}%
\bibitem [{\citenamefont {Dong}\ \emph {et~al.}(2009)\citenamefont {Dong},
  \citenamefont {Faessler}, \citenamefont {Gutsche}, \citenamefont
  {Kovalenko},\ and\ \citenamefont {Lyubovitskij}}]{dong2009PRD79-094013}%
  \BibitemOpen
  \bibfield  {author} {\bibinfo {author} {\bibfnamefont {Y.}~\bibnamefont
  {Dong}}, \bibinfo {author} {\bibfnamefont {A.}~\bibnamefont {Faessler}},
  \bibinfo {author} {\bibfnamefont {T.}~\bibnamefont {Gutsche}}, \bibinfo
  {author} {\bibfnamefont {S.}~\bibnamefont {Kovalenko}},\ and\ \bibinfo
  {author} {\bibfnamefont {V.~E.}\ \bibnamefont {Lyubovitskij}},\ }\bibfield
  {title} {\bibinfo {title} {${{X}}(3872)$ as a hadronic molecule and its
  decays to charmonium states and pions},\ }\href
  {https://doi.org/10.1103/PhysRevD.79.094013} {\bibfield  {journal} {\bibinfo
  {journal} {Phys. Rev. D}\ }\textbf {\bibinfo {volume} {79}},\ \bibinfo
  {pages} {094013} (\bibinfo {year} {2009})},\ \Eprint
  {https://arxiv.org/abs/0903.5416} {arxiv:0903.5416 [hep-ph]} \BibitemShut
  {NoStop}%
\bibitem [{\citenamefont {Fleming}\ and\ \citenamefont
  {Mehen}(2012)}]{fleming2012PRD85-014016}%
  \BibitemOpen
  \bibfield  {author} {\bibinfo {author} {\bibfnamefont {S.}~\bibnamefont
  {Fleming}}\ and\ \bibinfo {author} {\bibfnamefont {T.}~\bibnamefont
  {Mehen}},\ }\bibfield  {title} {\bibinfo {title} {The decay of the
  ${X}(3872)$ into $\chi_{cJ}$ and the operator product expansion in {XEFT}},\
  }\href {https://doi.org/10.1103/PhysRevD.85.014016} {\bibfield  {journal}
  {\bibinfo  {journal} {Phys. Rev. D}\ }\textbf {\bibinfo {volume} {85}},\
  \bibinfo {pages} {014016} (\bibinfo {year} {2012})},\ \Eprint
  {https://arxiv.org/abs/1110.0265} {arxiv:1110.0265 [hep-ph]} \BibitemShut
  {NoStop}%
\bibitem [{\citenamefont {Chen}(2020)}]{chen2020CPC44-023103}%
  \BibitemOpen
  \bibfield  {author} {\bibinfo {author} {\bibfnamefont {Y.-H.}\ \bibnamefont
  {Chen}},\ }\bibfield  {title} {\bibinfo {title} {Predictions of
  $\mathrm{\ensuremath{\Upsilon}}(4{S}) \to h_b(1{P},2{P}) \pi^+\pi^-$
  transitions},\ }\href {https://doi.org/10.1088/1674-1137/44/2/023103}
  {\bibfield  {journal} {\bibinfo  {journal} {Chin. Phys. C}\ }\textbf
  {\bibinfo {volume} {44}},\ \bibinfo {pages} {023103} (\bibinfo {year}
  {2020})},\ \Eprint {https://arxiv.org/abs/1907.05547} {arxiv:1907.05547
  [hep-ph]} \BibitemShut {NoStop}%
\bibitem [{\citenamefont {Chen}\ \emph {et~al.}(2017)\citenamefont {Chen},
  \citenamefont {Cleven}, \citenamefont {Daub}, \citenamefont {Guo},
  \citenamefont {Hanhart}, \citenamefont {Kubis}, \citenamefont {Mei{\ss}ner},\
  and\ \citenamefont {Zou}}]{chen2017PRD95-034022}%
  \BibitemOpen
  \bibfield  {author} {\bibinfo {author} {\bibfnamefont {Y.-H.}\ \bibnamefont
  {Chen}}, \bibinfo {author} {\bibfnamefont {M.}~\bibnamefont {Cleven}},
  \bibinfo {author} {\bibfnamefont {J.~T.}\ \bibnamefont {Daub}}, \bibinfo
  {author} {\bibfnamefont {F.-K.}\ \bibnamefont {Guo}}, \bibinfo {author}
  {\bibfnamefont {C.}~\bibnamefont {Hanhart}}, \bibinfo {author} {\bibfnamefont
  {B.}~\bibnamefont {Kubis}}, \bibinfo {author} {\bibfnamefont {U.-G.}\
  \bibnamefont {Mei{\ss}ner}},\ and\ \bibinfo {author} {\bibfnamefont {B.-S.}\
  \bibnamefont {Zou}},\ }\bibfield  {title} {\bibinfo {title} {Effects of
  ${Z}_{b}$ states and bottom meson loops on
  $\mathrm{\ensuremath{\Upsilon}}(4{S})\ensuremath{\rightarrow}\mathrm{\ensuremath{\Upsilon}}(1{S},2{S}){\ensuremath{\pi}}^{+}{\ensuremath{\pi}}^{\ensuremath{-}}$
  transitions},\ }\href {https://doi.org/10.1103/PhysRevD.95.034022} {\bibfield
   {journal} {\bibinfo  {journal} {Phys. Rev. D}\ }\textbf {\bibinfo {volume}
  {95}},\ \bibinfo {pages} {034022} (\bibinfo {year} {2017})},\ \Eprint
  {https://arxiv.org/abs/1611.00913} {arxiv:1611.00913 [hep-ph]} \BibitemShut
  {NoStop}%
\bibitem [{\citenamefont {Guo}\ \emph {et~al.}(2010)\citenamefont {Guo},
  \citenamefont {Hanhart}, \citenamefont {Li}, \citenamefont {Mei{\ss}ner},\
  and\ \citenamefont {Zhao}}]{guo2010PRD82-034025}%
  \BibitemOpen
  \bibfield  {author} {\bibinfo {author} {\bibfnamefont {F.-K.}\ \bibnamefont
  {Guo}}, \bibinfo {author} {\bibfnamefont {C.}~\bibnamefont {Hanhart}},
  \bibinfo {author} {\bibfnamefont {G.}~\bibnamefont {Li}}, \bibinfo {author}
  {\bibfnamefont {U.-G.}\ \bibnamefont {Mei{\ss}ner}},\ and\ \bibinfo {author}
  {\bibfnamefont {Q.}~\bibnamefont {Zhao}},\ }\bibfield  {title} {\bibinfo
  {title} {Novel analysis of the decays
  ${\ensuremath{\psi}}{\ensuremath{'}}\ensuremath{\rightarrow}{h}_{c}{\ensuremath{\pi}}^{0}$
  and
  ${\ensuremath{\eta}}_{c}{\ensuremath{'}}\ensuremath{\rightarrow}{\ensuremath{\chi}}_{c0}{\ensuremath{\pi}}^{0}$},\
  }\href {https://doi.org/10.1103/PhysRevD.82.034025} {\bibfield  {journal}
  {\bibinfo  {journal} {Phys. Rev. D}\ }\textbf {\bibinfo {volume} {82}},\
  \bibinfo {pages} {034025} (\bibinfo {year} {2010})},\ \Eprint
  {https://arxiv.org/abs/1002.2712} {arxiv:1002.2712 [hep-ph]} \BibitemShut
  {NoStop}%
\end{thebibliography}%
\end{document}